\documentclass[sigconf]{acmart}

\usepackage{tikz}
\usepackage{amsmath}
\usepackage{amsfonts} 
\usepackage{amsmath}
\usepackage{algorithm}
\usepackage{xcolor}
\usepackage{algpseudocode}
\algrenewcommand\algorithmiccomment[1]{\textit{\textcolor{gray}{// #1}}} %
\usepackage{subfigure}
\usepackage{gensymb}
\usepackage{multirow}
\usepackage{multicol}
\usepackage{booktabs}
\usepackage{soul}
\usepackage{subfiles}
\usepackage{colortbl}

\AtBeginDocument{%
  }

\copyrightyear{2025}
\acmYear{2025}
\setcopyright{acmlicensed}
\acmConference[CCS '25] {Proceedings of the 2025 ACM SIGSAC Conference on Computer and Communications Security}{ October 13--17, 2025}{Taipei, Taiwan.}
\acmBooktitle{Proceedings of the 2025 ACM SIGSAC Conference on Computer and Communications Security (CCS '25), October 13--17, 2025, Taipei, Taiwan}
\acmDOI{10.1145/3719027.3765092}
\acmISBN{979-8-4007-1525-9/2025/10}



\begin{document}


\title{Asymmetry Vulnerability and Physical Attacks on Online Map Construction for Autonomous Driving}

\author{Yang Lou*}
\affiliation{%
  \institution{City University of Hong Kong}
  \city{Hong Kong}
  \state{}
  \country{China}
}
\email{yanglou3-c@my.cityu.edu.hk}
\thanks{$^*$Equal contribution.}

\author{Haibo Hu*}
\affiliation{%
  \institution{City University of Hong Kong}
  \city{Hong Kong}
  \state{}
  \country{China}
}
\email{haibohu2-c@my.cityu.edu.hk}

\author{Qun Song*}
\affiliation{%
  \institution{City University of Hong Kong}
  \city{Hong Kong}
  \state{}
  \country{China}
}
\email{qunsong@cityu.edu.hk}

\author{Qian Xu}
\affiliation{%
  \institution{City University of Hong Kong\\ City University of Hong Kong Matter Science Research Institute (Futian)}
  \city{Hong Kong}
  \state{}
  \country{China}
}
\email{qian.xu@cityu.edu.hk}

\author{Yi Zhu}
\affiliation{%
  \institution{Wayne State University}
  \city{Detroit}
  \state{}
  \country{USA}
}
\email{yzhu39@wayne.edu}

\author{Rui Tan}
\affiliation{%
  \institution{Nanyang Technological University}
  \city{}
  \state{}
  \country{Singapore}
}
\email{tanrui@ntu.edu.sg}

\author{Wei-Bin Lee}
\affiliation{%
  \institution{Information Security Center, Hon Hai Research Institute \\ Feng Chia University}
  \city{Taipei, Taiwan}
  \state{}
  \country{}
}
\email{wei-bin.lee@foxconn.com}

\author{Jianping Wang$^\dagger$}
\affiliation{%
  \institution{City University of Hong Kong}
  \city{Hong Kong}
  \state{}
  \country{China}
}
\email{jianwang@cityu.edu.hk}
\thanks{$^\dagger$Corresponding author.}

\renewcommand{\shortauthors}{Yang et al.}
\newcommand{\red}{\color{red}}
\newcommand{\blue}{\color{blue}}
\renewcommand{\vec}[1]{\mathbf{#1}} %
\newcommand{\vecg}[1]{\boldsymbol{#1}} %

\begin{abstract}
High-definition (HD) maps provide precise environmental information essential for prediction and planning in autonomous driving (AD) systems. Due to the high cost of labeling and maintenance, recent research has turned to online HD map construction using onboard sensor data, offering wider coverage and more timely updates for autonomous vehicles (AVs). However, the robustness of online map construction under adversarial conditions remains underexplored.
In this paper, we present a systematic vulnerability analysis of online map construction models, which reveals that these models exhibit an inherent bias toward predicting symmetric road structures. In asymmetric scenes like forks or merges, this bias often causes the model to mistakenly predict a straight boundary that mirrors the opposite side. We demonstrate that this vulnerability persists in the real-world and can be reliably triggered by obstruction or targeted interference.
Leveraging this vulnerability, we propose a novel two-stage attack framework capable of manipulating online constructed maps. First, our method identifies vulnerable asymmetric scenes along the victim AV’s potential route. Then, we optimize the location and pattern of camera-blinding attacks and adversarial patch attacks. Evaluations on a public AD dataset demonstrate that our attacks can degrade mapping accuracy by up to 9.9\% in average precision, render up to 44\% of targeted routes unreachable, and increase unsafe planned trajectory rates—colliding with real-world road boundaries—by up to 27\%. These attacks are also validated on a real-world testbed vehicle.\footnote{Our real-world attack demos are available at \url{http://onlinemapattack.online/}.}
We further analyze root causes of the symmetry bias,  attributing them to training data imbalance, model architecture, and map element representation. Based on these findings, we propose asymmetric data fine-tuning as a targeted defense, which significantly improves model robustness.
To the best of our knowledge, this study presents the first vulnerability assessment of online map construction models and introduces the first digital and physical attack against them.

\end{abstract}

\maketitle


\section{Introduction}
\begin{figure}[t]
    \centering
    \includegraphics[width=0.9\columnwidth]{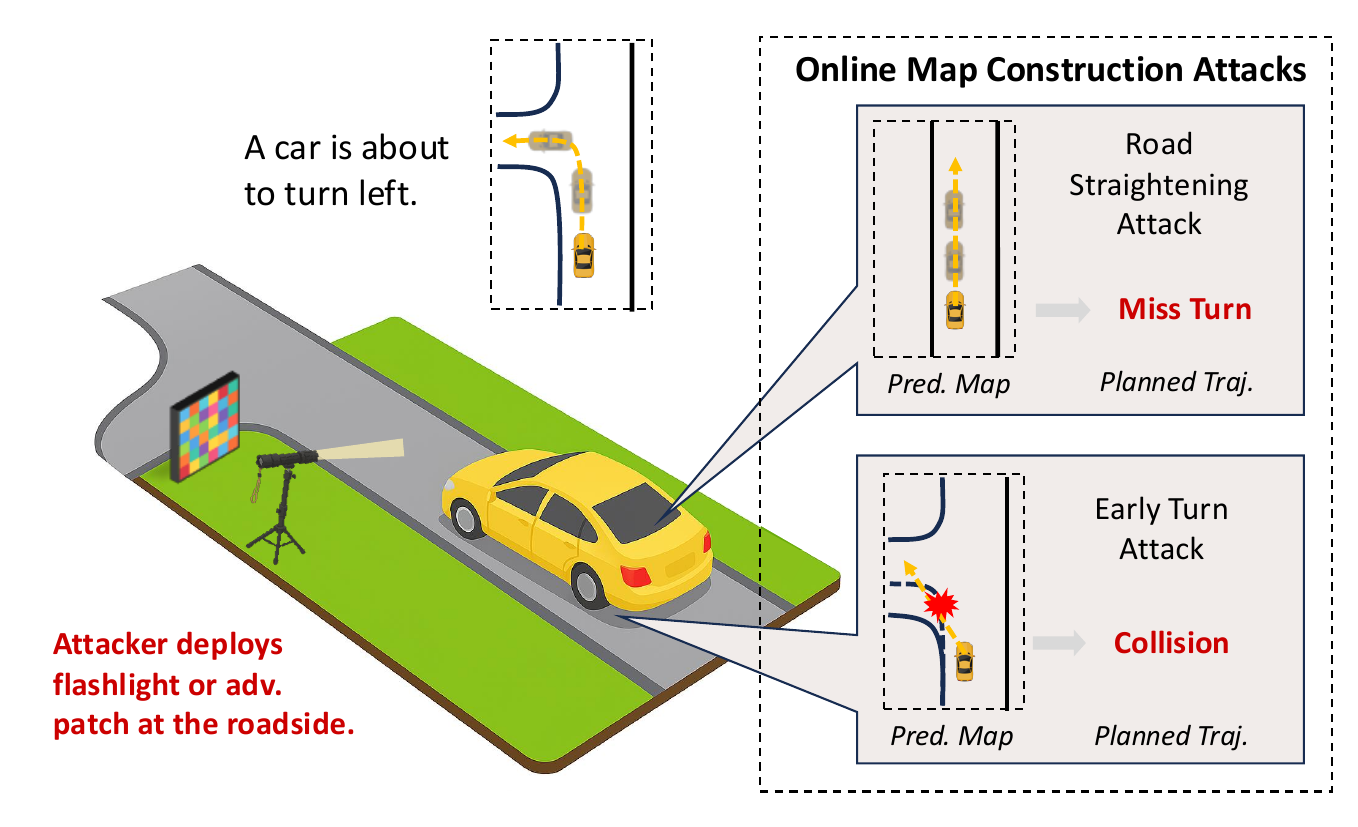}
    \vspace{-0.5em}
    \caption{By placing a flashlight or adversarial patch at the roadside of an asymmetric scene, an attacker can mislead the autonomous vehicle’s online map construction, potentially inducing unsafe planning decisions.}\label{fig:attack_scenario}
    \vspace{-1em}
\end{figure}

High-definition (HD) maps are essential for modern autonomous driving (AD) systems, providing precise and structured environmental representation for reliable decision-making. These maps capture centimeter-level details of vectorized map elements, along with traffic and navigation data. Traditional HD map construction relies on SLAM-based pipelines with manual annotation, which is labor-intensive. Recent advances focus on online HD map construction that predicts vectorized representation for road boundaries, lane dividers, and pedestrian crossings using onboard sensors, enabling fresher maps at lower costs. AD companies further scale this process by aggregating maps from fleet vehicles in a crowd-sourced fashion to build and maintain global HD maps.
HD maps support downstream modules like trajectory prediction and motion planning. Consequently, any inaccuracies can lead to unsafe driving behaviors such as missed turns, lane drifting, or collisions. Worse, if a flawed map is uploaded to a crowd-sourced global map, it can corrupt shared data and mislead all vehicles relying on it. 
Despite their critical role in AD pipelines, the security and robustness of online map construction remain underexplored.


Prior physical attacks targeting map-related data flow in AD systems mainly focus on lane detection, which uses a front-view image to identify lane markings and supports advanced driver assistance systems (ADAS) functions like lane-keeping and lane-changing. These attacks generally launch in two ways: (1) placing on-road objects, such as crafted patches~\cite{sato2021dirty, jing2021too} or traffic cones~\cite{han2022physical,zhang2024towards}; and (2) exploiting environmental conditions, such as projecting phantom lanes at night~\cite{nassi2020phantom} or leveraging sun-induced shadow patterns~\cite{zhang2024lanevil, mohajeransari2024discovering}. However, existing attacks suffer from practical limitations: on-road object placement often violates traffic regulations and is costly. Phantom-based attacks depend on specific settings like nighttime or sun position.
Unlike lane detection, which only generates the traffic lanes, online map construction generates vectorized maps, including pedestrian crossings, lane dividers, road boundaries, etc. The richer input, broader output range, and more complex model architecture of online map construction, which feature deep interaction between map elements and image features, introduce new challenges and attack opportunities. 
In this work, we propose a low-cost and physically realizable roadside attack that disrupts online map construction and cascades into consequential failures in downstream planning.
\textbf{Vulnerability Analysis and Attack Opportunity.} To understand whether online HD map construction models have exploitable weaknesses, we conduct a vulnerability analysis using a large-scale autonomous driving dataset. 
We classify scenes into two categories: \textit{symmetric scene}, where left and right road boundaries have similar or mirrored structures (e.g., straight roads and intersections); and \textit{asymmetric scene}, where one boundary significantly diverges while the other remains relatively straight (e.g., forks and merges). 
Our analysis shows that online map construction tends to reconstruct symmetric maps even when the ground truth environment is asymmetric. For example, the model often reconstructs a straight road map in scenarios where the actual environment is a road fork.
These results reveal an inherent bias in the online map construction models toward predicting symmetric road structures. 
We further validate this bias in real-world experiments using our testbed AV running an online map construction model. We evaluate one symmetric and one asymmetric scene under three conditions: (1) clean, (2) flashlight interference at a random roadside position, and (3) flashlight placed based on our attack framework. The symmetric scene remains robust across all conditions. In contrast, the asymmetric scene is correctly predicted in the clean case, mildly distorted under random interference, and completely mispredicted as a symmetrical straight road under our targeted attack, confirming that symmetry bias can be reliably triggered by physical interference with carefully chosen configurations.
This vulnerability creates an exploitable attack surface for physical attacks against online map construction. Specifically, we consider two physical attack vectors, flashlights and adversarial patches, that introduce physical interference to achieve two key objectives: (1) Road Straightening, which hides turns by inducing false straight-boundary predictions; and (2) Early Turn, which shifts predicted boundary earlier than intended, increasing the risk of roadside collisions.

\textbf{Challenges in Exploiting Symmetry Bias.} Effectively exploiting the symmetry bias in online map construction models presents three key challenges. 
First, while asymmetric scenes are more susceptible to attacks, they are not explicitly labeled in existing map datasets, making it difficult to identify these scenes without expert knowledge. This necessitates the development of an automated asymmetric scene detection approach, either for launching attacks or for large-scale robustness testing by AD companies.
Second, practical attack deployment faces substantial real-world constraints. Attacks should be deployed from the roadside following traffic regulations, with limited resources such as fixed-size adversarial patches or flashlights with constrained power. 
Third, the search space for attack configuration, including attack position, height, and patch pattern, is large and complex, particularly when targeting a victim's route or an entire urban area. Efficient optimization is needed to identify effective configurations within practical boundaries.

\textbf{Our Novel Attacks.} To address the above three challenges, we propose a two-stage attack framework. The first stage identifies vulnerable asymmetric scenes using HD/SD or self-constructed maps and optional camera images. A rule-based geometric classifier detects asymmetry by analyzing curvature differences between road boundaries, while a vision-language model (VLM) filters out false positives and adds semantic labels and reasoning for reference. In the second stage, we optimize physical attack configurations under real-world constraints. To reduce the search space, we design a scoring mechanism that ranks candidate roadside positions by balancing attack intensity and coverage of critical asymmetry regions. For execution, we define task-specific loss functions for two attack objectives: road straightening and early turn. Camera blinding attacks simulate lens flare effects and use black-box heuristic search to find effective flashlight positions. Adversarial patch attacks simulate projected patterns and apply a hybrid of heuristic search and projected gradient descent (PGD) to jointly optimize patch placement and appearance. This framework supports four possible attack configurations, derived by combining two attack vectors with two attack objectives, each capable of effectively manipulating online-constructed maps in asymmetric driving scenarios.

We evaluate our attack on both the nuScenes dataset and a custom-built testbed AV. In the dataset-based experiments, our attack reduces map construction average precision (AP) by up to 9.9\%, causes 44\% of scenes to result in unreachable route planning due to road straightening, and increases the unsafe planned trajectory rate-collisions with real-world road boundaries-to 27\% due to early turn predictions.
In real-world tests, our attack successfully induces symmetry prediction errors in five asymmetric scenes using either flashlight-based camera blinding or adversarial patches.
We identify three root causes of this vulnerability: training data imbalance, network design, and map element representation. Based on these insights, we propose asymmetric data fine-tuning, which improves model performance and reduces the attack's effectiveness on asymmetric scenes.

Our main contributions are summarized as follows:
\begin{itemize}
    \item We conduct the first systematic vulnerability analysis of online map construction models and reveal an inherent bias toward symmetric road predictions under obstructions or targeted interference.
    \item We propose a novel two-stage attack framework that efficiently identifies attack configurations for black-box camera blinding and white-box adversarial patches, enabling effective road straightening and early turn attacks.
    \item We evaluate our attack on both a public autonomous driving dataset and a real-world testbed, demonstrating significant degradation in map quality and planning outcomes, including unreachable routes and collisions. We further analyze root causes and propose a defense to mitigate the threat.
\end{itemize}

The remainder of this paper is organized as follows:
Section~\ref{sec:background} reviews related work.
Section~\ref{sec:vul} introduces the identified vulnerability.
Section~\ref{sec:threat_model} defines the threat model.
Section~\ref{sec:method} details our attack framework.
Sections~\ref{sec:exp_dataset} and~\ref{sec:exp_realworld} present dataset-based and real-world evaluations.
Section~\ref{sec:root_causes} analyzes root causes and proposes a defense.
Section~\ref{sec:limitations} discusses limitations and future directions.
Section~\ref{sec:conclusion} concludes the paper.

\section{Background}\label{sec:background}

\subsection{Autonomous Driving Systems} 
Autonomous driving (AD) systems typically consist of perception, trajectory prediction, and motion planning modules, each essential for safe and effective vehicle operation.
The \textit{perception} module processes sensor data from cameras, LiDAR, radar, and other onboard sensors to interpret the environment. Two primary perception tasks are (1) \textit{object detection and tracking}, and (2) \textit{mapping}. Object detection and tracking identify dynamic objects, such as vehicles, pedestrians, and cyclists, producing structured outputs in the form of bounding boxes and their trajectories over time. Mapping generates high-definition (HD) maps that provide detailed geometric and semantic information about the static environment, including road boundaries, lane dividers, and pedestrian crossings. 
\textit{Trajectory prediction} bridges perception and planning by forecasting future trajectories of surrounding objects based on perceived states and HD maps. Finally, the \textit{motion planning} module integrates perception and prediction outputs with HD map information to generate a safe, efficient, and feasible driving trajectory.
Given their critical influence on downstream decision-making, the reliability of HD maps is vital for AD system safety and robustness.

\vspace{-1em}
\subsection{Online Map Construction} 
Traditionally, HD maps are built offline using simultaneous localization and mapping (SLAM)-based methods, followed by labor-intensive manual annotation. This approach incurs high costs and produces maps that are outdated quickly, requiring frequent maintenance. To address these challenges, online map construction leverages real-time onboard sensor data to dynamically predict static map elements, enabling more efficient and up-to-date mapping. 
Consequently, this technique has been widely adopted by leading AD companies (e.g., MobilEye~\cite{mobileye_rem}, XPeng~\cite{xpeng_map_report} and Li Auto~\cite{liu2023vectormapnet}) and map vendors (e.g., TomTom~\cite{tomtom_orbis_adas} and Baidu Maps~\cite{xia2024dumapnet}).

Early online map construction methods employ a rasterized pipeline~\cite{li2022bevformer, philion2020lift, zhou2022cross}, transforming surround-view images into a unified Bird’s-Eye-View (BEV) representation and applying segmentation to generate semantic maps. However, rasterized maps lack structural consistency and instance-level clarity, limiting their utility in prediction and planning.
Recent advancements have shifted towards vectorized map prediction, which directly predicts structured map elements as polylines or polygons, providing a more compact and interpretable output. 
Formally, given a set of surround-view images $\mathcal{I} = \{I_1, I_2, \dots, I_K\}$, where each $I_k \in \mathbb{R}^{H \times W \times C}$, the task can be defined as learning a model:
$\mathcal{M}: \mathcal{I} \rightarrow \mathcal{V} = \{V_1, V_2, \dots, V_N\}$,
where each $V_i$ is defined as an ordered set of 2D points $V_i = \{v_{i,1}, v_{i,2}, \dots, v_{i,T_i}\}$, representing a polyline (e.g., road boundary, lane divider) or a polygon (e.g., pedestrian crossing) in BEV coordinates.
Methods like HDMapNet~\cite{li2022hdmapnet} and InstaGraM~\cite{shin2025instagram} use graph structures and vertex clustering for vectorized map prediction. 
More recent methods, including VectorMapNet~\cite{liu2023vectormapnet} and the MapTR series~\cite{MapTR, liao2024maptrv2}, proposed by Horizon Robotics and Li Auto, respectively, adopt query-based decoding for improved accuracy and efficiency. In our experiments, we employ both VectorMapNet and MapTR as representative industry-proposed solutions.
Beyond these, various works~\cite{xiong2023neural, chen2024mapcvv, qin2023traffic} explore the construction of a global map using the online constructed map in a crowdsourcing manner.
Despite rapid progress, robustness remains an open challenge. MapBench~\cite{hao2024your} evaluates existing models under adverse weather and sensor failures, revealing significant performance degradation. However, the vulnerability of these models to adversarial attacks remains largely unexplored, posing a critical risk to their safe and reliable real-world deployment.

While lane detection also addresses static scene understanding, it differs significantly in scope, input, and output. Lane detection typically operates on a single front-view image and outputs 2D/3D lane markings in perspective view for real-time ADAS. In contrast, online map construction fuses surround-view images to produce a vectorized map of the full driving environment, including road boundaries, lane dividers, and pedestrian crossings, that directly supports downstream modules such as trajectory prediction and planning. Moreover, online maps can be aggregated over time to form consistent global HD maps. As such, errors in online map construction can have more severe and far-reaching impacts on high-level AD systems.


\vspace{-1em}
\subsection{Physical Attacks against Map Elements} 


To date, no research has investigated physical adversarial attacks against online map construction systems. Existing map-related attacks primarily focus on lane detection manipulation. 
These lane detection attacks aim to compromise vehicle lane-keeping or lane-change functions by inducing incorrect lane perception that causes vehicles to deviate from safe trajectories. 
Existing attacks employ strategies like exploiting environmental illusions like shadows and tire marks~\cite{zhang2024lanevil}, using negative shadow patterns by obstructing sunlight~\cite{mohajeransari2024discovering}, or applying subtle road markings and crafted patches to trick perception models into false lane recognition~\cite{jing2021too}. 
Researchers have also demonstrated how backdoor attacks can compromise lane detection systems. In these attacks, adversaries poison training data to implant hidden triggers in models that can be activated by common objects (e.g., traffic cones), leading to severe lane misdetection~\cite{han2022physical,zhang2024towards}. Additionally, phantom lanes projected via drones or digital billboards can mislead lane detection systems into recognizing nonexistent road elements~\cite{nassi2020phantom}.
However, attacks targeting maps used by downstream prediction or planning modules have received relatively little attention. Prior work~\cite{Zheng_2023_WACV} perturbs rasterized or vectorized maps to mislead trajectory prediction models, but these attacks require access to backend data servers and are conducted digitally rather than in the physical world.

Despite the task and model differences between lane detection and online map construction, existing lane detection attacks typically involve either placing physical objects on the road (e.g., patches, stickers, or traffic cones) or inducing patterns (e.g., projected phantom lanes or sun shadows). Object-based attacks may violate traffic laws, are costly to craft, and are easily detected and removed. Shadow-based attacks rely heavily on specific time and weather conditions. 
In this work, we introduce a low-cost, physically realizable attack targeting online map construction via subtle roadside interference.

\section{A General Model-level Vulnerability}\label{sec:vul}

\begin{table}[!t]
\centering
\caption{Comparison of scene type counts classified using ground truth maps versus online-constructed maps (bold indicates misclassified counts).}
\label{tab:vul_cls_clean}
\vspace{-0.5em}
\resizebox{0.9\columnwidth}{!}{
    \begin{tabular}{@{}cc|ccc@{}}
    \toprule
    Model                  & GT Scene & Pred: Sym.  & Pred: Asym  & No Boundary \\ \midrule
    \multirow{2}{*}{MapTR} & Sym.     & 80          & \textbf{19} & 1         \\
                           & Asym.    & \textbf{31} & 69          & 0         \\ \midrule
    \multirow{2}{*}{\begin{tabular}[c]{@{}c@{}}Vector\\ MapNet\end{tabular}} & Sym. & 89 & \textbf{7} & 4 \\
                           & Asym.    & \textbf{60} & 32          & 8         \\ \bottomrule
    \end{tabular}
}
\end{table}

\subsection{Vulnerability Analysis}\label{sec:vul_analysis}
Our investigation of MapTR~\cite{MapTR} and VectorMapNet~\cite{liu2023vectormapnet} predictions on the nuScenes dataset~\cite{caesar2020nuscenes} reveals a significant pattern: 
both models perform well on straight roads and symmetrical intersections.
However, they struggle with asymmetric scenarios. Specifically, in fork scenarios, these models often fail to accurately predict turning boundaries and instead predict them as straight lines, indicating a fundamental model-level vulnerability. 
To systematically analyze this behavior, we categorize scenes based on the structural relationship between left and right road boundaries:
\begin{itemize}
    \item \textbf{Symmetric scenes:} Both left and right road boundaries exhibit similar or mirrored structures, such as straight roads with approximately parallel edges or intersections diverging symmetrically.
    \item \textbf{Asymmetric scenes:} One road boundary significantly diverges (e.g., makes a turn or splits off) while the opposite boundary remains relatively straight. Common examples include road forks and lane merging.
\end{itemize}

\subsubsection{Vulnerability Analysis in Dataset.} 
To quantify model performance, we select 100 \textit{symmetric} scenes and 100 \textit{asymmetric} scenes from the nuScenes validation set, classified using the ground truth annotation maps as described in Section~\ref{sec:method_identify_asym}.
We evaluate how online map construction models including VectorMapNet~\cite{liu2023vectormapnet} and MapTR~\cite{MapTR} predict vectorized maps from six surround-view camera images. Scene types are then re-classified based on the predicted maps using the same classification method.
Table~\ref{tab:vul_cls_clean} compares ground truth and predicted scene types. 
For example, in the \textit{MapTR} row under \textit{Sym.} GT scenes, the \textit{Pred: Sym.} column shows that $80$ ground-truth symmetric scenes (e.g., straight roads, intersections) are correctly predicted as symmetric, reflecting accurate map predictions with minimal impact on downstream planning. 
The bold entries highlight misclassified scenes caused by map prediction errors. For MapTR, $19$ out of $100$ symmetric scenes were misclassified as asymmetric. More notably, $35$ out of $100$ asymmetric scenes were misclassified as symmetric without any interference.
These results suggest that \textbf{online map construction models possess an inherent bias toward predicting symmetric road structures.}

\begin{figure}[!t]
    \centering
    \hfill
    \subfigure[Asym. Scene.]{
        \label{fig:asym_scene_cam}
        \includegraphics[width=0.28\columnwidth]{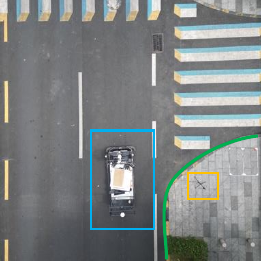}
    }
    \hfill
    \subfigure[Random Interference.]{
        \label{fig:asym_scene_random}
        \includegraphics[width=0.28\columnwidth]{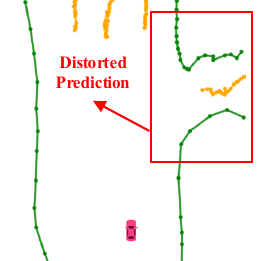}
    }\hfill
    \subfigure[Our Attack.]{
        \label{fig:asym_scene_attack}
        \includegraphics[width=0.28\columnwidth]{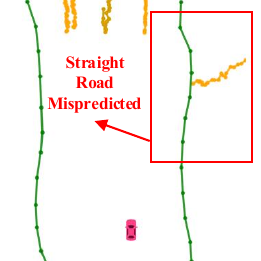}
    }\hfill
    \\
    \hfill
    \subfigure[Sym. Scene]{
        \label{fig:sym_scene_cam}
        \includegraphics[width=0.28\columnwidth]{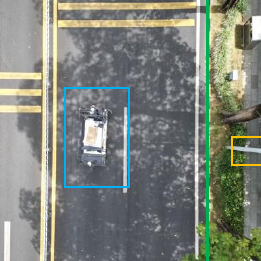}
    }
    \hfill
    \subfigure[Random Interference.]{
        \label{fig:sym_scene_random}
        \includegraphics[width=0.28\columnwidth]{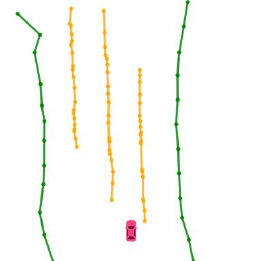}
    }\hfill
    \subfigure[Our Attack.]{
        \label{fig:sym_scene_attack}
        \includegraphics[width=0.28\columnwidth]{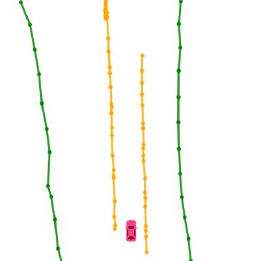}
    }\hfill
    \vspace{-0.5em}
    \caption{Real-world vulnerability experiments in symmetric and asymmetric scenes under clean conditions, random interference, and our attack (victim AV: blue box; flashlight: orange box).}
    \vspace{-1em}
\end{figure}

\subsubsection{Vulnerability Analysis in Real-World}
To validate this vulnerability under physical conditions, we conduct real-world experiments using our customized testbed car running MapTR. We select one symmetric and one asymmetric scene and compare the model’s online map predictions under three conditions: (1) clean (no interference), (2) random flashlight interference, and (3) flashlight positioned based on our attack framework. The two scenarios are illustrated in Fig.~\ref{fig:asym_scene_cam} and Fig.~\ref{fig:sym_scene_cam}.
In the clean setting, the model correctly predicts a straight road in the symmetric scene and a right turn in the asymmetric scene. With random interference, predictions remain stable in the symmetric case (Fig.~\ref{fig:sym_scene_random}) and only slightly distorted in the asymmetric case (Fig.~\ref{fig:asym_scene_random}), preserving the turn.
However, when the flashlight is placed at the position optimized by our attack framework, the model mispredicts a straight road in the asymmetric scene (Fig.~\ref{fig:asym_scene_attack}), suppressing the right turn, while the symmetric scene remains unaffected (Fig.~\ref{fig:sym_scene_attack}).
These results demonstrate that \textbf{the symmetry bias can lead to real-world threats when triggered via physical interference, particularly with carefully chosen attack configurations.}

\begin{figure}[!t]
    \centering
    \hfill
    \subfigure[Road straightening.]{
        \label{fig:vul_attack_oppo_straightening}
        \includegraphics[width=0.4\columnwidth]{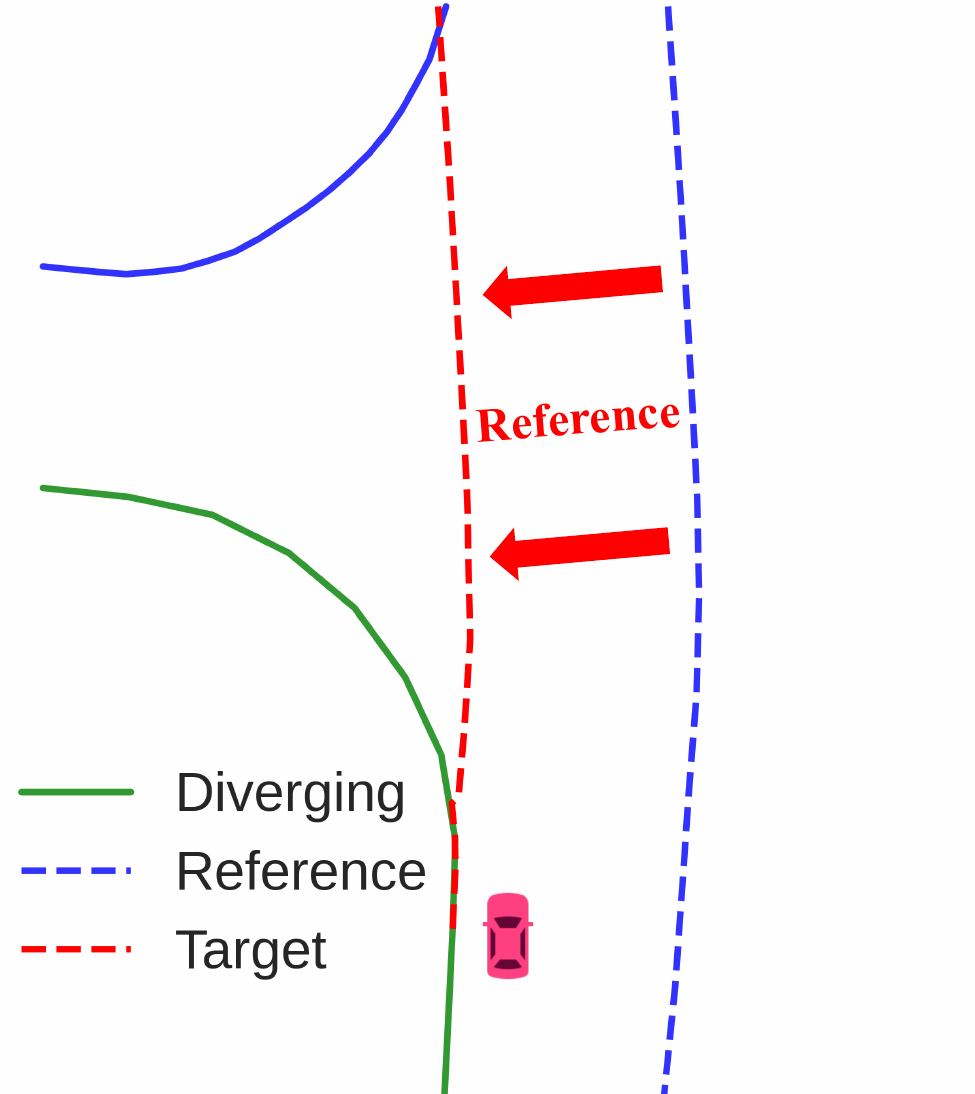}
    }\hfill
    \subfigure[Early turn.]{
        \label{fig:vul_attack_oppo_earlyturn}
        \includegraphics[width=0.4\columnwidth]{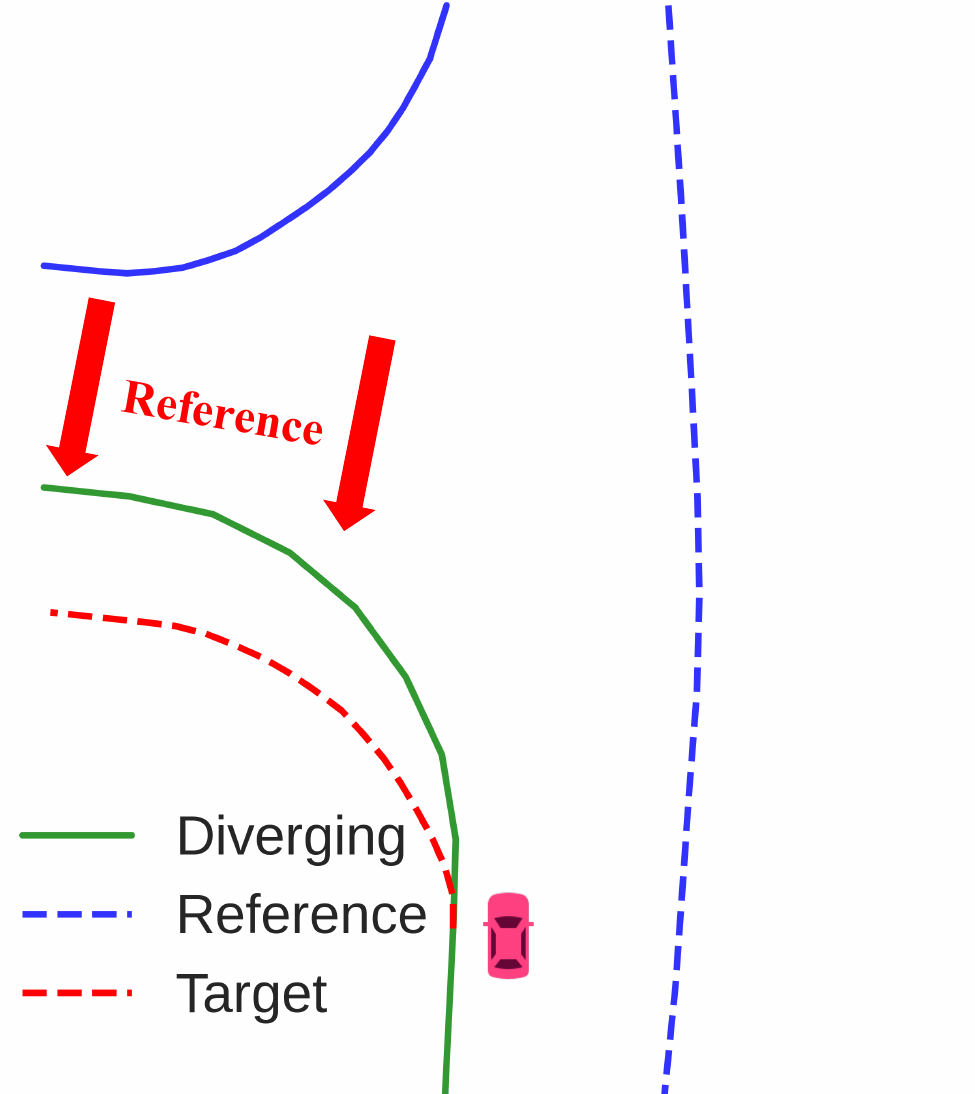}
    }\hfill
    \vspace{-0.5em}
    \caption{Attack Opportunities. The green curve indicates the diverging (left-turn) boundary. Dashed blue lines show reference boundaries: the right-side boundary before the fork enables road straightening attacks, while the forward boundary after the turn enables early turn attacks.}\label{fig:vul_attack_oppo}
    \vspace{-1em}
\end{figure}

\subsection{Attack Opportunity from Symmetry Bias}
To understand why specific attack configurations trigger symmetry bias, we analyze the architecture of online map construction models. These models, including MapTR, typically employ a BEV encoder–map decoder architecture. The BEV encoder extracts contextual features from surround-view images, while the map decoder uses attention mechanisms to model interactions between map elements. As a result, map element predictions are heavily influenced by both visual context and nearby elements. 

In symmetric scenes, mutual boundary references help produce stable and accurate predictions. 
However, in asymmetric scenes, this becomes a vulnerability. Even under clean conditions, the model may incorrectly predict symmetry when left and right boundaries differ, as shown in our dataset analysis.
When physical interference is applied at a carefully chosen position, such as obscuring the right turn in Fig.~\ref{fig:asym_scene_cam}, critical visual context is disrupted. The model then relies more on nearby elements, leading to incorrect symmetry cues from the opposite boundary and ultimately an incorrect symmetric prediction.
Exploiting this bias in asymmetric scenes is significantly more effective and efficient than directly manipulating geometry in symmetric ones.

To formalize the attack strategy, we define three key roles in asymmetric road structures: The \textit{Attack Target Boundary} is the predicted boundary targeted by the attack. The \textit{Diverging Boundary} is the ground truth boundary of the attack target boundary. The \textit{Reference Boundary} provides contextual cues used to mislead its prediction. 
We also define the \textit{Asymmetry Anchor} as the position where the diverging boundary begins to deviate, typically at turns, forks, or merges. It serves as a prior for selecting effective attack configurations.
The model’s symmetry bias creates two critical attack opportunities: (1) \textit{Road Straightening} (Fig.~\ref{fig:vul_attack_oppo_straightening}), which suppresses the diverging turn (green) by encouraging the model to mirror the straight reference boundary (dashed blue) on the right, leading to an incorrectly straightened prediction (dashed red), and (2) \textit{Early Turn} (Fig.~\ref{fig:vul_attack_oppo_earlyturn}), which induces early boundary shifts (dashed red) by referencing the reference boundary (dashed blue).

To achieve these objectives, we propose two attack vectors involving physical interference: (1) \textit{Camera Blinding}, which uses a directed flashlight to temporarily obscure critical positions in camera views; and (2) \textit{Adversarial Patches} strategically placed roadside elements that induce targeted interference when viewed by cameras.
Both vectors support practical roadside deployment and exploit models' symmetry bias in asymmetric scenes.

\section{Threat Model}\label{sec:threat_model}

\subsection{Attack Goal.}\label{sec:attack_goal}
We consider the attack scenario shown in Fig.~\ref{fig:attack_scenario}, where an attacker places a flashlight or adversarial patch at the roadside of an asymmetric scene. A victim AV approaches, relying on an online map construction model to guide its motion planning.

Given limited attack resources, we aim to exploit the online map construction model's symmetry bias to mislead the model into predicting a symmetric road structure in asymmetric environments, potentially causing dangerous driving behavior.
Formally, the goal is to identify an effective attack configuration $ \theta^*$ that transforms the surround-view images into $\mathcal{I}' = \mathcal{T}(\mathcal{I}, \theta)$, such that the predicted diverging boundary $ \mathcal{M}(\mathcal{I}')_{div}$ aligns with a target symmetric boundary $ \mathcal{V}_{tar}$:
\begin{equation}\label{eq:problem_def}
    \theta^* = \arg\min_{\theta} \mathcal{L}(\mathcal{M}(\mathcal{I}')_{div}, \mathcal{V}_{tar}), 
\end{equation}
where $\mathcal{L}$ is an objective-specific loss that measures the alignment between the predicted and the target boundary. The formulations of $\mathcal{L}$ and $\mathcal{V}_{tar}$ are detailed in Section~\ref{sec:method_attack_optimization} on Attack Design.
We consider two specific attack objectives:
\begin{itemize}
    \item \textbf{Road Straightening Attack (RSA)} misleads the model into omitting diverging paths, making turns unreachable.
    \item \textbf{Early Turn Attack (ETA)} causes the predicted road boundary to turn earlier, potentially leading to collisions with actual road edges.
\end{itemize}
To achieve these objectives, the attack configuration for camera blinding is the flashlight position, constrained by the brightness of commercial flashlights. For the adversarial patch attack, the attack configuration includes the patch's position and pattern, limited by practical patch size constraints.
 
These attacks have significant real-world implications. 
Road straightening attacks can create unreachable routes. An attacker may exploit this for economic gain, such as by blocking access to a business through the removal of a parking lot entrance.
Early turn attacks, especially when collision-driven, may involve intentional harm or unethical competition aimed at undermining the safety reputation of rival autonomous driving companies.
Both types of attacks, when triggered mid-turn, can cause sudden braking or sudden lane changes, potentially leading to rear-end collisions or traffic jams—endangering passengers, surrounding vehicles, and other road users.
Moreover, corrupted online-constructed maps could be uploaded to the cloud and propagated, compromising global maps and misleading all users relying on them.
At scale, automated attacks across urban areas could pose widespread safety and mobility risks. 
On the constructive side, these vulnerabilities also highlight scenarios that AV companies can use for targeted testing and robustness improvement, ultimately enhancing system safety.
Notably, our attack can affect not only a specific victim AV but also other AVs that pass the same location with a line of sight to the deployed attack vectors.

\vspace{-0.5em}
\subsection{Attacker's Knowledge and Capabilities.}
\textbf{Offline Attack.} We consider a realistic offline setting where the attacker has no real-time access to the victim AV. Instead, the attacker uses an SD/HD map or a pre-constructed map of the AV’s potential route or target area to generate optimized attack configurations.

\textbf{Black-box vs. White-box Settings.} 
We consider both the black-box and white-box settings to launch the attacks. 
In the black-box setting (applicable to camera blinding attacks), the attacker can query the online map construction model but lacks access to internal gradients.
In the white-box setting (applicable to adversarial patch attacks), the attacker has full access to the model, including its architecture and gradients.

\textbf{Deployment Constraints.} The attacker is limited to roadside deployment rather than on-road placement within predefined scenarios, to avoid possible detection or removal. 
For the camera blinding attack, a flashlight or projector is positioned at the roadside to temporarily obstruct the vehicle’s cameras. For adversarial patch attacks, patches can be affixed to roadside boards or displayed on digital billboards. To further minimize the risk of detection or removal by road maintenance or surveillance systems, the flashlight or projector can be activated only when the target victim AV is approaching, and the adversarial patch can be embedded in selected frames of a video shown on a roadside digital billboard or projected using a movable drone.

\section{Attack Design}\label{sec:method}
\begin{figure*}[!t]
    \centering
    \includegraphics[width=\textwidth]{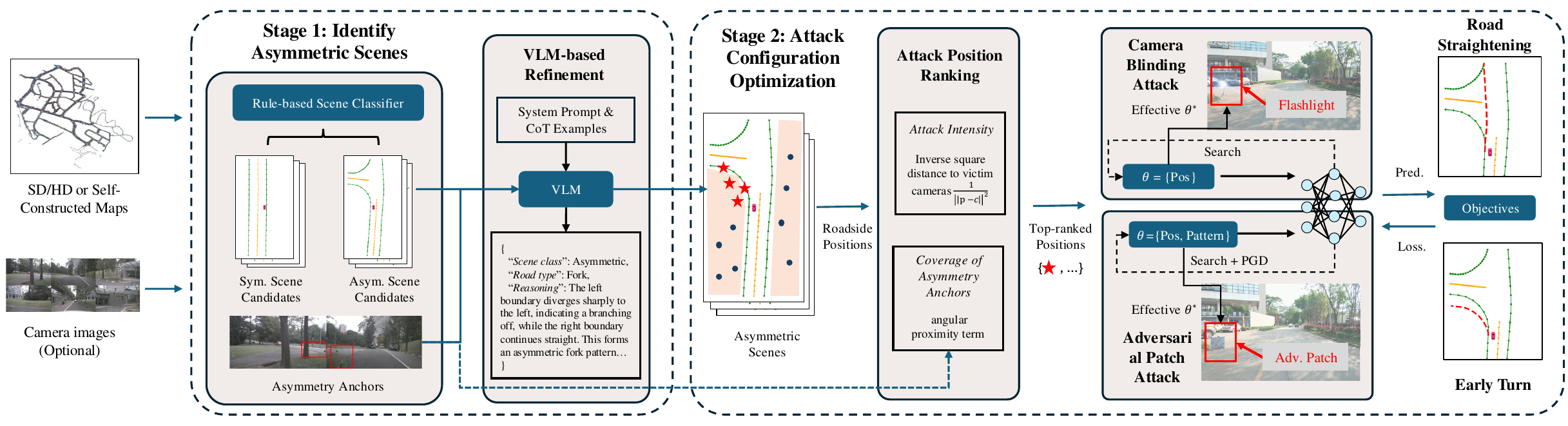}
    \vspace{-1.5em}
    \caption{Overview of our two-stage attack framework for identifying effective configurations to launch camera blinding and adversarial patch attacks.}\label{fig:attack_framework}
\end{figure*}

We design a two-stage attack framework, illustrated in Fig.~\ref{fig:attack_framework}, that identifies attack configurations for configuring flashlights or adversarial patches to effectively mislead the victim AV into predicting symmetric maps in asymmetric environments, ultimately inducing hazardous driving behaviors.

\subsection{Attack Challenges}\label{sec:challenges}
Our vulnerability analysis reveals several significant challenges in developing effective real-world attacks against online map construction models.

\textbf{Challenge C1: Automatic Detection of Attack-Prone Scenes.}\label{challenge_asym}
Our vulnerability analysis reveals that asymmetric scenes show degraded performance and are more susceptible to attack interference. To exploit this, attackers or AD developers conducting robustness testing must first identify such scenes along the victim vehicle’s route or across broader urban areas.
However, existing map data neither label scenes as symmetric or asymmetric, nor indicate where asymmetry begins. SD and HD maps lack explicit annotations of boundary symmetry or asymmetry anchors, which are critical for targeting vulnerable regions. Similarly, self-constructed maps lack such information.
Manually identifying asymmetric scenes and asymmetry anchors is labor-intensive and requires expert knowledge. An automated method is essential to support large-scale vulnerability assessment, enabling both attackers and AD companies to detect and analyze high-risk asymmetric scenarios without expert intervention efficiently.

\textbf{Challenge C2: Vast Search Space for Attack Configurations.}\label{sec:challenge_efficient}
Crafting an effective attack necessitates precisely determining where and how to deploy the interference.
For camera blinding attacks, this includes optimal roadside positions and heights of the flashlight.
For adversarial patches, the configuration space expands to include visual patterns and deployment angles.
Since the victim's route may only be determined at run time and the attack configurations are high-dimensional, the search space becomes extraordinarily vast. When targeting a global map in a large urban area, the computational cost of basic searching methods, such as brute-force search, becomes prohibitively high. 
An efficient strategy is therefore essential to reduce the search space and identify effective configurations under resource and time constraints.

\textbf{Challenge C3: Effective and Practical Attacks against Online Map Construction Models.}\label{sec:challenge_effective}
Manipulating an online constructed map is significantly more challenging than targeting objects or lane detections. 
First, map elements like road boundaries span large physical areas and appear across multiple camera views. Second, map construction models leverage sophisticated contextual reasoning and model interactions between map elements. 
These features necessitate broader and more strategic physical interference to successfully alter map predictions.
However, practical and legal constraints restrict attackers in our attack scenario to deploy flashlights or adversarial patches only from the roadside positions. This increases the distance to the victim AV, which reduces attack visibility and weakens attack effectiveness, creating a trade-off between attack effectiveness and real-world feasibility. 
These challenges demand an attack optimization framework capable of identifying effective configurations within practical deployment constraints.

\vspace{-1em}
\subsection{Design Overview}
Our attack framework consists of two stages: (1) asymmetric scene identification and (2) attack configuration optimization.

\textbf{Stage 1: Asymmetric Scene Identification.}
To address Challenge \textbf{C1} and automatically identify asymmetric scenes vulnerable to attack, we introduce a two-step classification method in Section~\ref{sec:method_identify_asym}. In the first step, a rule-based geometric classifier analyzes SD/HD or self-constructed maps to identify scenes with significant left-right curvature differences and locate asymmetry anchors. In the second step, a vision-language model (VLM) filters false positives and incorporates semantic reasoning. This ensures accurate detection of vulnerable asymmetric scenes and provides critical guidance for the following attack process.

\textbf{Stage 2: Attack Configuration Optimization.}
This stage addresses both efficiency (Challenge \textbf{C2}) and effectiveness (Challenge \textbf{C3}). We first reduce the vast configuration space by ranking candidate roadside positions using a lightweight scoring function (Section~\ref{sec:method_position_ranking}) based on attack intensity and coverage of critical asymmetry anchors. The top-ranked positions form a finite attack position set for further optimization.
Next, we optimize attack configurations (Section~\ref{sec:method_attack_optimization}) by simulating the visual effects of flashlights and adversarial patches. We define two map manipulation objectives: road straightening and early turn. For the black-box camera blinding attack, we apply heuristic search over positions; for the white-box adversarial patch, we use a hybrid strategy, searching positions and applying PGD for pattern optimization.
This framework produces four effective attack configurations, combining two physical vectors with two objectives, each capable of disrupting online HD map construction in asymmetric driving scenarios.

\vspace{-0.5em}
\subsection{Identify Asymmetric Scenario}\label{sec:method_identify_asym}
Asymmetric scenes represent key vulnerabilities in online map construction. To support targeted attacks and robustness evaluation, we use a lightweight two-step classification pipeline that takes map data and optional camera images as input and outputs a set of identified asymmetric scenes along with their corresponding asymmetry anchor set $\mathcal{D}$. These results serve as inputs to the subsequent attack configuration optimization stage.

\subsubsection{Step 1: Rule-based Classification}

In this step, we employ geometric analysis to detect asymmetry by quantifying curvature differences between left and right road boundaries. 
This approach is motivated by the observation that asymmetric scenes typically exhibit distinct geometric behaviors between the left and right boundaries, particularly in scenarios where one boundary significantly curves while the other remains relatively straight. 
Formally, given left and right boundaries represented as ordered sets of 2D points in BEV coordinates, $V_\text{left} = \{v_{l,1}, v_{l,2}, \dots, v_{l,T_l}\}$ and $V_\text{right} = \{v_{r,1}, v_{r,2}, \dots, v_{r,T_r}\}$, where each point $v_{i,j} = (x_{i,j}, y_{i,j})$, we calculate the point-wise curvature $k$ at each point using the general plane curve parametrization: 
$k = \frac{|x^{\prime} y^{\prime\prime} - y^{\prime} x^{\prime\prime}|}{(x^{\prime2} + y^{\prime2})^{3/2}}.$
Here, $x^{\prime}, y^{\prime}$ and $x^{\prime\prime}, y^{\prime\prime}$ represent the first and second derivatives of the boundary coordinates, respectively. For robustness, we compute regional curvature $\bar{k}$ by averaging curvature values $k$ within sliding windows along each boundary. The curvature difference ($\Delta k$) is defined as:
\begin{equation}
\begin{split}
    \Delta k &= \max(|k_\text{left}(t) - k_\text{right}(t)|), \\
    \text{s.t.} \quad &\min(\bar{k}_\text{left}(t), \bar{k}_\text{right}(t)) < \bar{k}_{\text{thre}},
\end{split}
\end{equation}
where $\bar{k}_\text{left}(t)$ and $\bar{k}_\text{right}(t)$ denote regional curvature values at corresponding positions along the boundaries, and $\bar{k}_{\text{thre}}$ refers to an empirically determined curvature threshold. A scene is identified as asymmetric if the curvature difference exceeds the threshold ($\Delta k > \Delta k_{\text{thre}}$) while maintaining relatively low curvature on one of the boundaries. 

During this analysis, we also identify critical asymmetry anchors along the road boundaries to identify positions where significant geometric deviations occur. These points are defined as positions along the diverging boundary where the curvature difference exceeds a predefined threshold. We denote this set as $\mathcal{D}=\{d_1, \dots, d_K\}$, where each $d_i$ represents a 2D BEV coordinate. These anchors provide key cues for VLM-based refinement and guide attack position scoring in the subsequent optimization framework.

\subsubsection{Step 2: VLM-based Refinement}

While the rule-based geometric approach in step 1 effectively identifies many asymmetric scenarios, real-world road structures often exhibit complex, irregular geometries that cause false positives. For example, a symmetric crossroad may be misclassified as asymmetric due to irregular boundary shapes, resulting in high curvature differences (see Fig.~\ref{fig:vlm_refinement_map} in the Appendix).
To address this, we introduce a second classification step utilizing a vision-language model (VLM), which offer expert-level road structure and driving scene understanding. We avoid directly applying the VLM to all scenes where it may misclassify without strong contextual cues;  
instead, the rule-based method provides coarse filtering and locates asymmetric anchors, which serve as visual cues that guide the VLM in handling difficult corner cases with improved accuracy.

\textbf{VLM Inputs.} For each scene initially classified as asymmetric, we construct a comprehensive multi-modal input for the VLM, combining structured textual map data with visualized map and camera cues. 
Each input includes: (1) structured JSON data containing boundary coordinates and positions with significant curvature differences, marked as potential asymmetry anchors; (2) a BEV map visualization showing left and right boundaries, the ego vehicle, and asymmetry anchors; and (3) front-view camera images with red bounding boxes highlighting asymmetric anchors. This multi-modal representation provides both map-level and driver-perspective context to support accurate VLM reasoning.

\textbf{System Prompt and CoT Reasoning.} The VLM is guided by a structured system prompt that defines its role, required skills, classification criteria, task description, and input/output formats. To improve interpretability and accuracy, the prompt incorporates a chain-of-thought (CoT) reasoning strategy, instructing the model to analyze the road layout, compare boundary behavior, integrate visual and map cues, make a classification, and assess potential safety risks. We also provide three illustrative examples, covering both symmetric and asymmetric cases with expected outputs. Full prompt details are available in Appendix Fig.~\ref{fig:vlm_system_prompt}.

\textbf{VLM Outputs and Refinement.} The VLM outputs a structured JSON response containing the final classification (symmetric/asymmetric), specific the road type when applicable (e.g., fork, turn, lane merging, etc.), and detailed reasoning supporting its decision. We use this output to refine the initial geometric classifications, keeping only scenes confirmed as asymmetric by the VLM. Fig.~\ref{fig:vlm_refinement_output} in the Appendix shows a successful correction, where the VLM (GPT-4o) correctly classifies a previously misidentified crossroad as symmetric, despite irregular left-boundary shapes.

This two-step classification pipeline effectively identifies genuine asymmetric road scenarios and their corresponding asymmetry anchors, forming the foundation of the attack framework.

\subsection{Attack Position Ranking} \label{sec:method_position_ranking}
Given the identified vulnerable asymmetric scenes in Section~\ref{sec:method_identify_asym}, we need to determine the most effective flashlight position for camera blinding attacks, or the most effective position–pattern pair for adversarial patch attacks. However, as noted in Challenge \textbf{C2}, directly optimizing over the entire 3D roadside space is computationally expensive and physically impractical. To address this, we introduce a lightweight ranking mechanism that narrows the search space by scoring candidate positions based on their geometric proximity and coverage to critical asymmetry anchors in the scene.
We observe that attacks are most effective when deployed near road boundary asymmetry anchors and in close proximity to the victim vehicle, where their influence on vision-based models is stronger. Guided by these insights, we define a scoring function $S(p)$ for each candidate position $p$, which generalizes across different attack vectors:
\begin{equation}
    S(p) = \sum_{c \in C} \sum_{d \in D} \begin{cases}
        \left(1 - \frac{\phi_{c,p,d}}{\phi_{\text{max}}}\right) \cdot \frac{1}{||p - c||^2}, & \text{if } \phi_{c,p,d} < \phi_{\max}, \\
        0, & \text{otherwise},
    \end{cases}
\end{equation}
where $C$ is the set of camera positions on the victim vehicle, $D$ is the set of critical asymmetry anchors, $\phi_{c,p,d}$ is the angle between the vectors from camera $c$ to position $p$, and from $c$ to asymmetry anchor $d$. $\phi_{\text{max}}$ represents the maximum influence angle, encompassing both the flashlight's beam angle and the effective range of the adversarial patch, and $||p - c||^2$ is the squared distance from the candidate position to the camera.
This formulation balances two key factors: attack intensity, modeled by the inverse-square distance term $\frac{1}{||p - c||^2}$, and coverage effectiveness, captured by the angular proximity term $\left(1 - \frac{\phi_{c,p,d}}{\phi_{\text{max}}}\right)$. The intensity term approximates the diminishing perceptual effect of lens flare or patch size with distance, while the coverage term favors positions aligned with the camera’s line of sight to critical asymmetry anchors. By combining these terms, the score $S(p)$ efficiently prioritizes positions that both maximize intensity and the interference of key road features.
We denote top-ranked position candidates selected by this scoring function as $\mathcal{P} = \{p_1, \dots, p_N\}$, where each $p_i \in \mathbb{R}^3$ represents a 3D roadside position. These ranked candidates serve as input to our full attack configuration optimization described next.

\subsection{Attack Configuration Optimization}
\label{sec:method_attack_optimization}

Given the ranked candidate positions $\mathcal{P}$, we now introduce an attack optimization framework to identify the most effective configuration $\theta^*$ and address Challenge \textbf{C3}. We detail the simulation methods for both camera blinding and adversarial patch attacks, define two strategic attack objectives with corresponding target boundary generation mechanisms, and present optimization strategies for each attack type. The framework outputs four optimized attack configurations $\theta^*$, covering two attack vectors and two objectives.

\begin{figure}[!t]
    \centering
    \includegraphics[width=0.85\columnwidth]{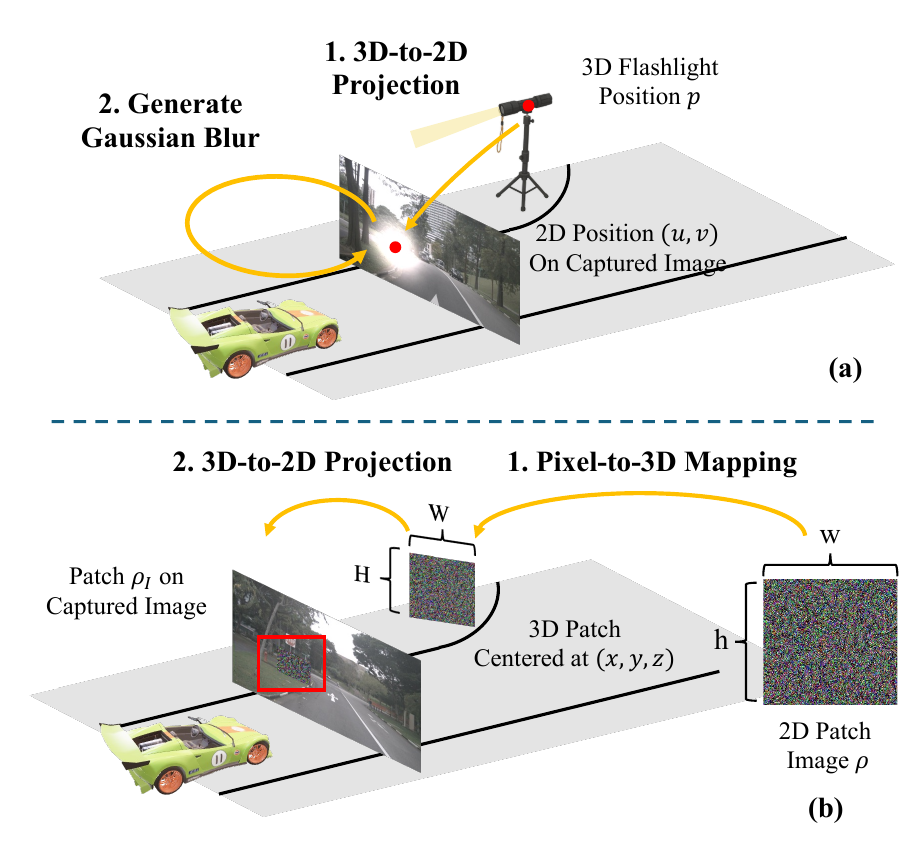}
    \vspace{-0.5em}
    \caption{Simulation of (a) camera blinding attack and (b) adversarial patch attack.}
    \label{fig:attack_simulation}
    \vspace{-1.4em}
\end{figure}

\subsubsection{Attack Simulation}

\textbf{Camera Blinding Simulation} models the visual interference caused by a high-intensity flashlight placed at a roadside location $p \in \mathcal{P}$, aiming to obscure critical input regions and mislead the map construction model. We optimize only the 3D position $\theta = p \in \mathbb{R}^3$ while keeping the physical parameters (e.g., intensity, beam angle, color temperature) fixed according to standard flashlight specifications.
Following~\cite{petit2015remote,zhang2020detecting}, we simulate the lens flare effect caused by the flashlight through two steps: 3D-to-2D projection and Gaussian blur generation, as shown in Fig.~\ref{fig:attack_simulation}(a).
In the first step, given the 3D position $p = (x^w, y^w, z^w)$ of the flashlight in the 3D world coordinate system, the process first transforms $p$ into the camera coordinate system and then projects it onto the 2D image plane of each camera view to obtain the coordinates $(u, v)$, which serve as the center of the blur. Based on the pinhole camera model, the transformation is formalized as:
\begin{equation}\label{eq:coord_trans}
\begin{bmatrix}
u \\ v \\ 1
\end{bmatrix}
=
\begin{bmatrix}
f_x & 0   & c_x \\
0   & f_y & c_y \\
0   & 0   & 1
\end{bmatrix}
\begin{bmatrix}
r_{11} & r_{12} & r_{13} & t_{11} \\
r_{21} & r_{22} & r_{23} & t_{12} \\
r_{31} & r_{32} & r_{33} & t_{13}
\end{bmatrix}
\begin{bmatrix}
x^w \\ y^w \\ z^w \\ 1
\end{bmatrix}
= K [R | T] W
\end{equation}
where $R$ and $T$ are the rotation matrix and translation vector that define the camera’s extrinsic parameters, and $(f_x, f_y, c_x, c_y)$ are the focal lengths and principal point offsets comprising the intrinsics $K$. All these parameters can be obtained through camera calibration.
In the second step, for each camera view with a valid projection $(u, v)$, we generate a Gaussian blur centered at that location. To simulate light intensity, we follow the inverse-square law: $L \propto 1/d^2$, where $d$ is the Euclidean distance between the flashlight position $p$ and the camera position in 3D world coordinate system. The blur radius is computed using logarithmic falloff to approximate realistic spatial light dispersion. 
In summary, the rendering process defines the transformation $\mathcal{T}$, which modifies the original surround-view input $\mathcal{I}$ to produce the blinded images $\mathcal{I}' = \mathcal{T}(\mathcal{I}, p)$.

\textbf{Adversarial Patch Simulation} places a physical patch at a candidate position $p \in \mathcal{P}$ with a learnable pattern $\rho \in \mathbb{R}^{h \times w \times 3}$. The physical dimensions (width and height) of the patch are predetermined based on practical deployment constraints such as resource limitations and concealment requirements. 
As shown in Fig.~\ref{fig:attack_simulation}(b), we follow the approach of~\cite{cheng2024fusion} to apply the adversarial patch to surround-view images, formalized as:
\begin{equation}
\begin{split}
    &\mathcal{I}' = \mathcal{I} \odot (1 - M_I) + \rho_{I} \odot M_I \\
    &M_I=proj_I(M), \rho_{I}=proj_I(\rho)\\
    &\rho\in[0, 1]^{3\times h\times w},M\in \{0, 1\}^{1\times h\times w}
\end{split}
\end{equation}
where $\rho$ and $M$ denote the initial patch image and binary mask, respectively, both defined in the 2D pixel coordinate system with width $w$ and height $h$.
To project the patch onto each camera view, the function $\text{proj}_I(\cdot)$ consists of two stages: pixel-to-3D mapping and 3D-to-2D projection.
In the first stage, each pixel $(u^p, v^p)$ in the patch image is mapped to 3D world coordinates $(x^w, y^w, z^w)$ as shown in Eq.~\ref{eq:pixel2world}. The patch's 3D position is specified by its center $p = (x^c, y^c, z^c)$, along with its width $W$, height $H$, and rotation angle $\alpha$. 
\begin{equation}\label{eq:pixel2world}
    \begin{bmatrix}
    x^{w} \\ y^{w} \\ z^{w} \\ 1
    \end{bmatrix}
    =
    \begin{bmatrix}
    \cos\alpha & 0 & -\sin\alpha & x^{c} \\
    0 & 1 & 0 & y^{c} \\
    \sin\alpha & 0 & \cos\alpha & z^{c} \\
    0 & 0 & 0 & 1
    \end{bmatrix}
    \;
    \begin{bmatrix}
    \frac{W}{w} & 0 & -\frac{W}{2} \\
    0 & \frac{H}{h} & -\frac{H}{2} \\
    0 & 0 & 0 \\
    0 & 0 & 1
    \end{bmatrix}
    \;
    \begin{bmatrix}
    u^{p} \\ v^{p} \\ 1
    \end{bmatrix}
\end{equation} 
In the second step, the corresponding 3D point is projected onto the image plane of each camera view using the perspective transformation in Eq.~\ref{eq:coord_trans}, resulting in the 2D coordinate $(u, v)$. This defines a quadrilateral region in the image representing the projected patch, denoted as $\rho_I$ and $M_I$.
This rendering process ensures that the patch appears physically consistent and visually coherent across all viewpoints.


\subsubsection{Attack Objectives}

\textbf{Road Straightening Objective}
aims to alter a diverging road boundary $\mathcal{V}_{div}$ into a straight boundary, effectively removing the road turns. To achieve this, we generate a target boundary $\mathcal{V}_{\text{tar}}$ by mirroring the geometry of a reference boundary $\mathcal{V}_{\text{ref}}$, producing a symmetric road structure.
Specifically, we first compute the average road width $w_{\text{avg}}$ between the diverging boundary $\mathcal{V}_{div}$ and the reference boundary $\mathcal{V}_{\text{ref}}$ near the vehicle. Using this width, we construct the target boundary as:
\begin{equation}
    \mathcal{V}_{tar} = \{v_{d,i} \mid i \leq k\} \cup \{(x_r \pm w_{\text{avg}}, y_r) \mid (x_r, y_r) \in V_{\text{ref}}\},
\end{equation}
where $k$ denotes the index of the diverging point marking the transition point between the straight segment and the diverging portion of the road. The sign of the shift ($+$ or $-$) is selected based on the relative positions of $\mathcal{V}_{div}$ and $V_{\text{ref}}$, ensuring that the target boundary lies on the side originally occupied by $\mathcal{V}_{div}$.
Fig.~\ref{fig:vul_attack_oppo_straightening} illustrates this artificially straightened boundary in a red dashed line. 
The optimization objective for this attack is defined as:
\begin{equation}
    \theta^* = \arg\min_{\theta} \mathcal{L}_{\text{chamfer}}\left(\mathcal{M}(\mathcal{T}(\mathcal{I}, \theta)), \mathcal{V}_{tar}\right),
\end{equation}
where $\mathcal{L}_{\text{chamfer}}$ is the Chamfer distance, measuring the average closest-point distance between the predicted and the target boundary. Minimizing this loss encourages the model to predict road boundaries that align with the artificially constructed symmetric structure, thereby inducing the desired road straightening effect.

\textbf{Early-Turn Objective}
aims to shift the predicted diverging road boundary outward toward the roadside, misleading the victim AV into initiating an early turn, which may lead to hazardous scenarios such as roadside collisions.
To achieve this goal, we design a directional loss function that explicitly encourages outward shifts of the boundary (toward the roadside) and penalizes inward shifts (toward the drivable area). Given the ground truth diverging boundary $\mathcal{V}_{div} = \{v_{d, i}\}$ and the centerline of the adjacent lane $C = \{c_i\}$, available from either HD/SD maps or self-collected data, we define outward direction vectors as: $d_i = \frac{v_{d, i} - c_i}{\|v_{d, i} - c_i\|}$. These unit vectors point from each centerline point toward the corresponding diverging boundary point and represent the desired direction of boundary displacement (as illustrated by red arrows in Fig.~\ref{fig:earlyturn_loss} in Appendix). Next, we compute the directional offset of each predicted boundary point $v_{d,i}' \in \mathcal{V}_{div}'$ from its ground truth position, projected along the outward direction: $\text{offset}_i = (v_{d,i}' - v_{d,i}) \cdot d_i$. The directional loss $\mathcal{L}_{\text{dir}}$ is then defined as:
\begin{equation}
    \mathcal{L}_{\text{dir}}(\mathcal{V}_{div}', \mathcal{V}_{div}, C) = \alpha \cdot \mathcal{L}_{\text{outward}} + \beta \cdot \mathcal{L}_{\text{inward}},
\end{equation}
where $\mathcal{L}_{\text{outward}} = -\text{ReLU}(\text{offset}_i)$ rewards outward displacement toward the roadside and $\mathcal{L}_{\text{inward}} = \text{ReLU}(-\text{offset}_i)$ penalizes inward displacement toward the center of the road, and $\alpha, \beta$ are scalar weights that control the balance between the two components. 
Finally, the optimization objective for the attack is expressed as:
\begin{equation}
    \theta^* = \arg\min_{\theta} \mathcal{L}_{\text{dir}}\left(\mathcal{M}(\mathcal{T}(\mathcal{I}, \theta)),\ \mathcal{V}_{div},\ C\right).
\end{equation}
Minimizing this loss encourages the predicted road boundary to shift outward in the direction of the roadside, thereby triggering early-turn behaviors in the victim AV.

\begin{table*}[!t]
\centering
\caption{Map AP(\%) on asymmetric scenes under our road straightening attacks.}
\label{tab:exp_rsa_map}
\vspace{-0.5em}
\begin{tabular}{@{}c|cccc|cccc@{}}
\toprule
\multirow{2}{*}{Method} & \multicolumn{4}{c|}{Blinding (Black-box)} & \multicolumn{4}{c}{Adv Patch (White-box)} \\
 & $AP_{boundary}$ & $AP_{divider}$ & $AP_{ped}$ & $mAP$ & $AP_{boundary}$ & $AP_{divider}$ & $AP_{ped}$ & $mAP$ \\ \midrule
Clean & 48.9 & 54.2 & 38.2 & 47.1 & 48.9 & 54.2 & 38.2 & 47.1 \\
Random Sampling & 43.3 & 48.4 & \textbf{34.9} & 42.2 & 44.7 & 50.9 & 38.5 & 44.7 \\
PSO & 42.4 & 48.9 & 37.3 & 42.8 & 44.8 & 51.6 & \textbf{37.1} & 44.5 \\
RSA (Ours) & \textbf{39.9} & \textbf{44.4} & 36.4 & \textbf{40.2} & \textbf{39.0} & \textbf{49.0} & 37.6 & \textbf{41.9} \\ \bottomrule
\end{tabular}
\end{table*}

\begin{table}[!t]
\centering
\caption{Unreachable Goal Rate (\%) on asymmetric scenes under our road straightening attacks.}
\label{tab:exp_rsa_plan}
\vspace{-0.5em}
\resizebox{0.95\columnwidth}{!}{
\begin{tabular}{@{}c|cc@{}}
\toprule
Method & Blinding (Black-box) & Adv Patch (White-box) \\ \midrule
Clean & \multicolumn{2}{c}{27} \\
Random Sampling & 34 (+7) & 33 (+6) \\
PSO & 37 (+10) & 34 (+7) \\
RSA (Ours) & \textbf{44 (+17)} & \textbf{44 (+17)} \\ \bottomrule
\end{tabular}
}
\vspace{-0.6em}
\end{table}


\subsubsection{Optimization Strategy}
We consider four attack configurations by combining two attack vectors (black-box camera blinding and white-box adversarial patch) with two attack objectives (road straightening and early turn). 
Our optimization strategy is tailored to each attack vector:
\begin{itemize}
    \item \textbf{Black-box camera blinding:} Due to the non-differentiable rendering process and discrete candidate positions, we perform a heuristic search over all $p \in \mathcal{P}$. We evaluate the objective-specific loss at each position and select the one with the lowest loss as the final configuration: $\theta^* = p^*$.
    \item \textbf{White-box adversarial patch:} We adopt a hybrid strategy combining heuristic search with Projected Gradient Descent (PGD). For each candidate position $p \in \mathcal{P}$, we optimize the patch pattern $\rho$ using PGD, leveraging the differentiable rendering process. The position–pattern pair with the lowest loss is selected as the final configuration: $\theta^* = \{p^*, \rho^*\}$.
\end{itemize}
The resulting configuration $\theta^*$ is then used to launch the attack.

\begin{figure}[!t]
    \centering
    \hfill
    \subfigure[Camera Blinding.]{
        \label{fig:exp_rsa_blind}   
        \includegraphics[width=0.36\columnwidth]{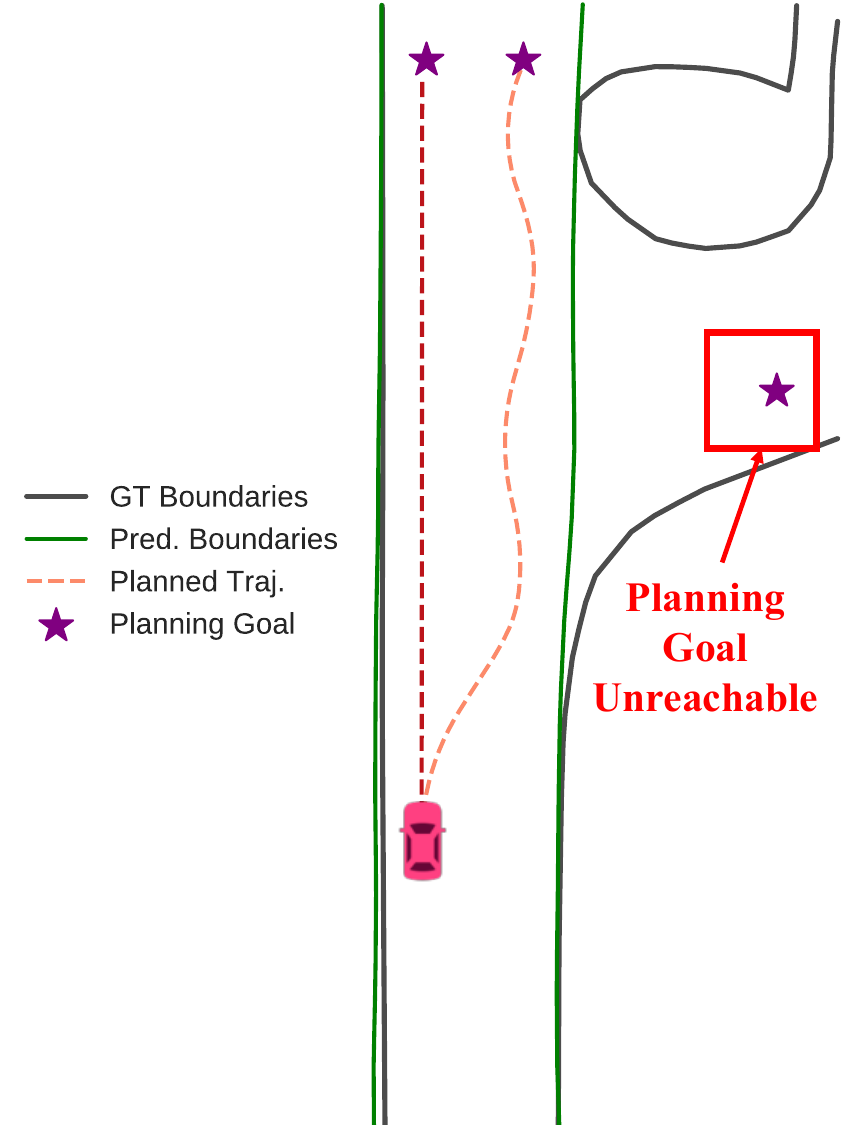}
    }\hfill
    \subfigure[Adversarial Patch.]{
        \label{fig:exp_rsa_patch}
        \includegraphics[width=0.36\columnwidth]{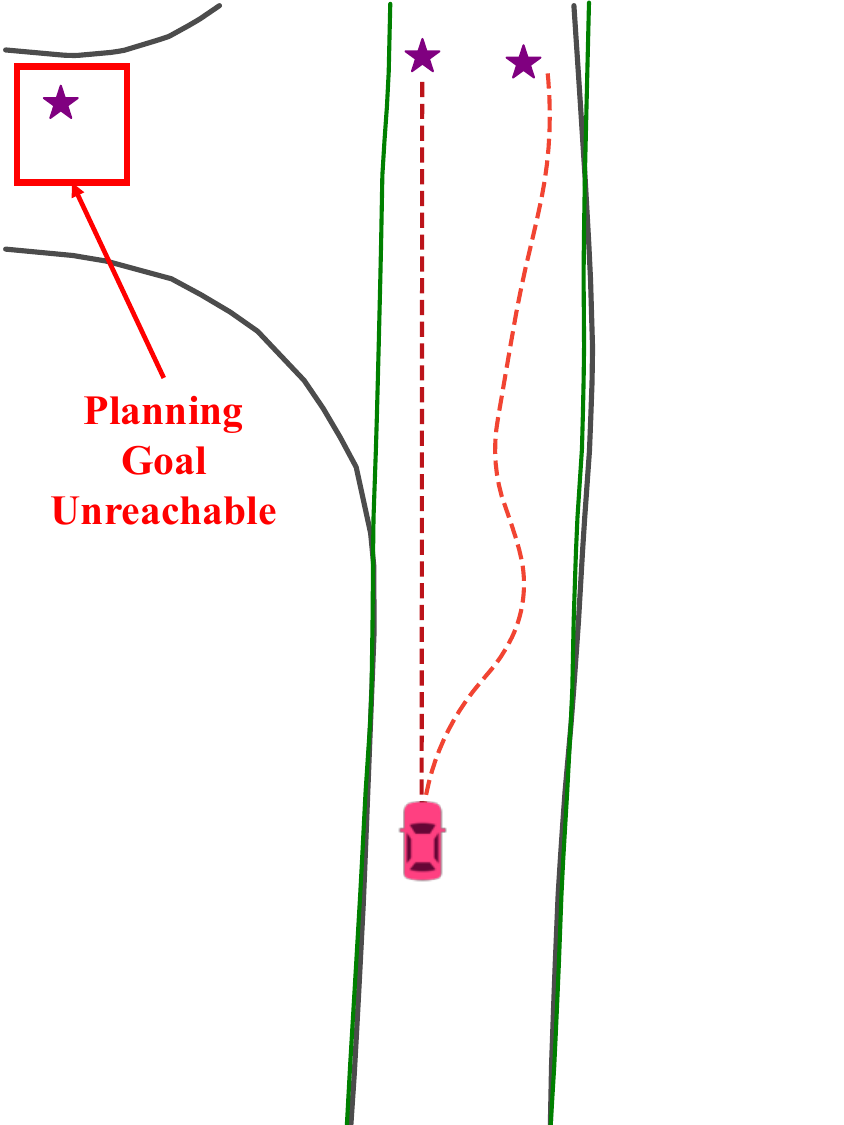}
    }\hfill
    \vspace{-0.8em}
    \caption{Examples of Road Straightening Attacks using camera blinding and adversarial patch.}\label{fig:exp_rsa_example}
    \vspace{-0.5em}
\end{figure}

\section{Experiments on Dataset}\label{sec:exp_dataset}
\begin{table*}[!t]
\centering
\caption{Map AP (\%) on asymmetric scenes under our early turn attacks.}
\label{tab:exp_eta_map}
\vspace{-0.7em}
\begin{tabular}{@{}c|cccc|cccc@{}}
\toprule
\multirow{2}{*}{Method} & \multicolumn{4}{c|}{Blinding (Black-box)} & \multicolumn{4}{c}{Adv Patch (White-box)} \\
 & $AP_{boundary}$ & $AP_{divider}$ & $AP_{ped}$ & $mAP$ & $AP_{boundary}$ & $AP_{divider}$ & $AP_{ped}$ & $mAP$ \\ \midrule
Clean & 48.9 & 54.2 & 38.2 & 47.1 & 48.9 & 54.2 & 38.2 & 47.1 \\
Random Sampling & 47.3 & 53.2 & 35.7 & 45.4 & 46.6 & 52.9 & 37.7 & 45.7 \\
PSO & \textbf{45.9} & \textbf{52.2} & 37.6 & 45.2 & 46.8 & 52.3 & \textbf{37.6} & 45.6 \\
ETA (Ours) & 46.2 & 52.5 & \textbf{34.5} & \textbf{44.4} & \textbf{44.2} & \textbf{51.1} & 38.3 & \textbf{44.5} \\ \bottomrule
\end{tabular}
\end{table*}

\begin{table}[!t]
\centering
\caption{Unsafe Planned Trajectory Rate (\%) on asymmetric scenes under early turn attacks.}
\label{tab:exp_eta_plan}
\vspace{-0.5em}
\resizebox{0.95\columnwidth}{!}{
\begin{tabular}{@{}c|cc@{}}
\toprule
Method & Blinding (Black-box) & Adv Patch (White-box) \\ \midrule
Clean & \multicolumn{2}{c}{10} \\
Random Sampling & 22 (+12) & 11 (+1) \\
PSO & 21 (+11) & 14 (+4) \\
ETA (Ours) & \textbf{27 (+17)} & \textbf{21 (+11)} \\ \bottomrule
\end{tabular}
}
\end{table}

In this section, we evaluate the effectiveness of our attack framework on asymmetric scenes from a public autonomous driving dataset. For comparisons across symmetric, asymmetric, and random scenes, see the Appendix~\ref{sec:attack_compare_sym_asym_rand}.

\vspace{-0.5em}
\subsection{Experiment Setting}
\textbf{Dataset.} We select 100 asymmetric driving frames from the nuScenes dataset~\cite{caesar2020nuscenes}, a large-scale real-world autonomous driving dataset with map annotations. Since the map annotations are incomplete in some regions (e.g., intersections), we first filter the validation set to include only frames with complete boundary data. We further exclude frames where the vehicle’s road lacks a left or right boundary, or where either boundary is shorter than 10 meters, as these are not meaningful for planning. This yields 2,095 valid frames. We then apply the method from Section~\ref{sec:method_identify_asym} to identify asymmetric frames. In the rule-based step, we set the curvature difference threshold to 0.3. For refinement, we use GPT-4o as the VLM analyzer. This process identifies 407 asymmetric frames, from which we randomly sample 100 for evaluation.

\textbf{Models.} We use MapTR~\cite{MapTR} for online map construction and the Hybrid A*-planner~\cite{sakai2018pythonrobotics} for motion planning in the victim AV’s system. MapTR is an industry-proposed, widely adopted model for online mapping. The planner directly uses the map generated by MapTR to plan trajectories. We train MapTR and implement the Hybrid A*-planner using their default settings.

\textbf{Flashlight and Adversarial Patch.} 
For the camera blinding attack, we use a flashlight with 3,000 lumens and a 40-degree beam angle, reflecting the upper limit of commercially available devices. For the adversarial patch attack, we set the patch size to $3m \times 2m$, approximating a roadside billboard, to ensure visibility even when the victim AV is centered on a wide road. In practice, patch size can be adjusted; for example, we use a smaller $1\mathrm{m} \times 1\mathrm{m}$ patch in the real-world experiment, which still proves effective.
In each evaluation scene, the attacker is allowed to deploy at most one flashlight or one patch at the roadside.

\textbf{Evaluation Metrics.} We use Average Precision (AP) to assess map construction accuracy, and Unreachable Goal Rate (UGR) and Unsafe Planned Trajectory Rate (UPTR) to measure planning impact under Road Straightening and Early Turn Attacks, respectively.

\textit{Average Precision (AP)}: The standard metric for online map construction, computed by averaging precision across multiple Chamfer Distance thresholds. We report \textit{$AP_{boundary}$}, \textit{$AP_{divider}$}, \textit{$AP_{ped}$} for road boundary, lane divider, and pedestrian crossing classes. The mean across all map elements is reported as \textit{mAP}.

\textit{Unreachable Goal Rate (UGR)}: The proportion of frames where the AV fails to generate a valid trajectory to one or more goal points. Higher UGR indicates greater route blockage due to mapping errors.

\textit{Unsafe Planned Trajectory Rate (UPTR)}: The proportion of frames where any planned trajectory intersects the ground-truth road boundary. A higher UPTR reflects an increased risk of near-future unsafe off-road behavior caused by map inaccuracies.

\textbf{Baseline: Random Sampling and PSO.} 
Since no prior attacks target online map construction, we use random sampling and Particle Swarm Optimization (PSO)~\cite{kennedy1995particle} as baseline methods for attack configuration selection, following their use in previous attacks~\cite{jing2021too,yu2025enduring}. 
These settings assume no knowledge of model vulnerabilities or prior information such as asymmetry anchors. Comparing them with our method under the same query budget highlights the efficiency of our Attack Position Ranking and the effectiveness of Attack Configuration Optimization.
For the camera blinding attack, the random sampling samples multiple roadside positions, while the PSO uses a swarm of particles to explore the search space within roadside regions; both methods then select the position that best achieves the attack objective.
For the adversarial patch attack, we evaluate two strategies: (1) Random Pattern: Sample multiple positions using random sampling or PSO and apply patches with random patterns; (2) Optimized Pattern: Sample multiple positions using random sampling or PSO and optimize the patch pattern using PGD at each selected position. For both strategies, we maintain the same total number of model queries to ensure fair comparison. We report the best-performing result from both strategies as the baseline in our experiments.

\begin{figure}[!t]
    \centering
    \hfill
    \subfigure[Camera Blinding.]{
        \label{fig:exp_eta_blind}
        \includegraphics[width=0.36\columnwidth]{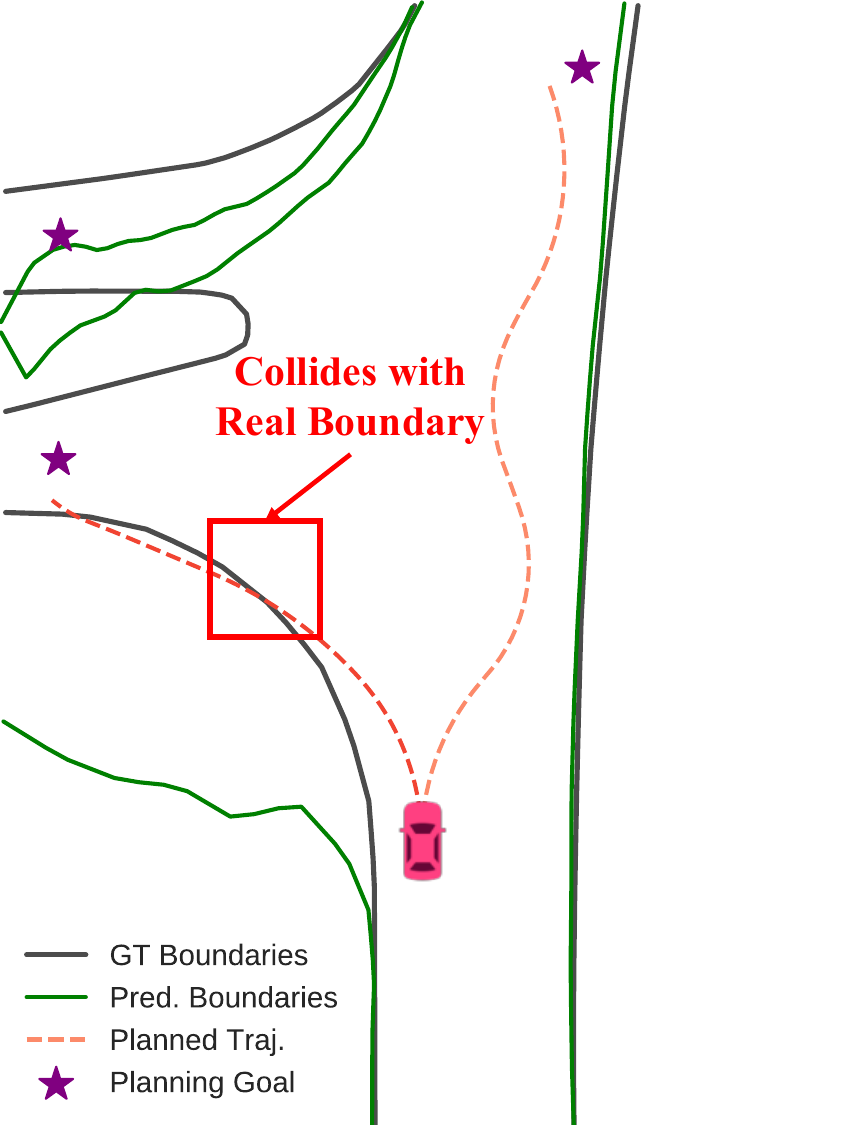}
    }\hfill
    \subfigure[Adversarial Patch.]{
        \label{fig:exp_eta_patch}
        \includegraphics[width=0.36\columnwidth]{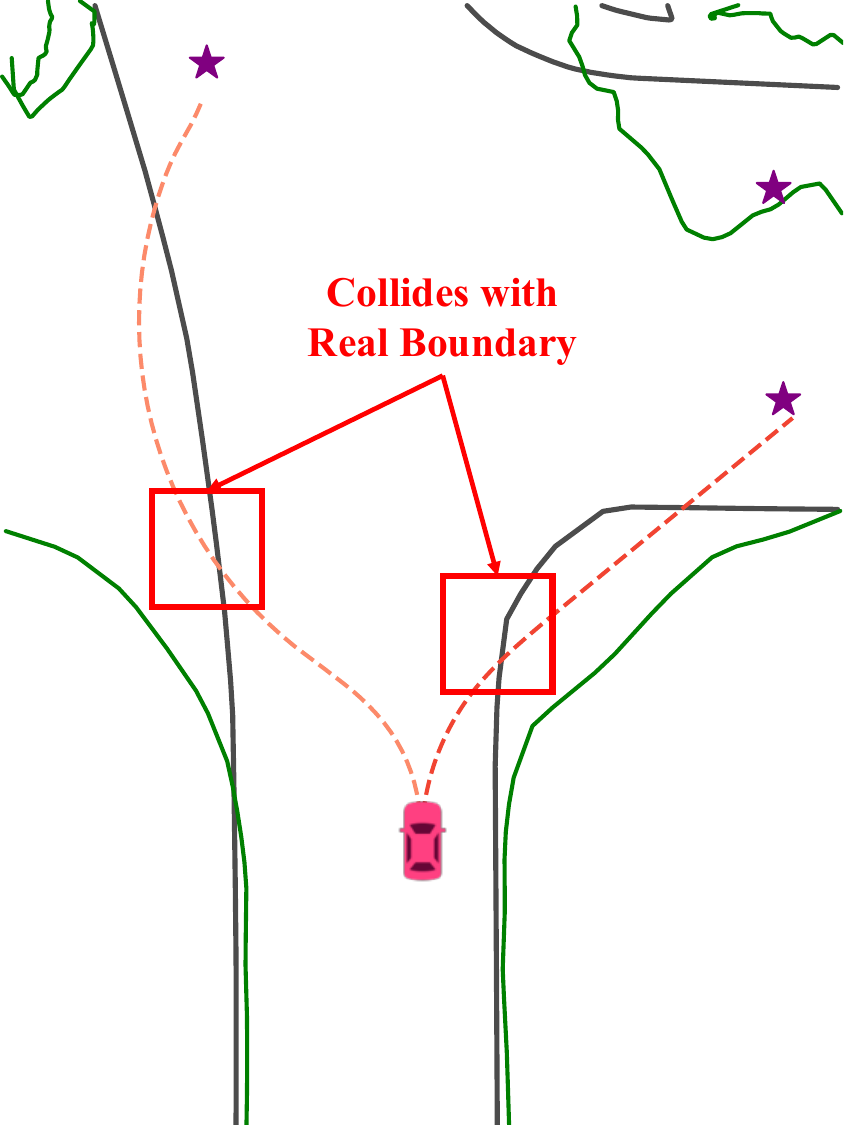}
    }\hfill
    \vspace{-1em}
    \caption{Examples of Early Turn Attacks using camera blinding and adversarial patches.}\label{fig:exp_eta_example}
\end{figure}

\subsection{Attack Effectiveness}
For fair comparisons, we evaluate attack effectiveness using AP, UGR, and UPTR under a fixed budget of 400 queries. This means the attacker can query the online map construction model up to 400 times. In adversarial patch attacks, total queries are calculated as the number of candidate positions multiplied by the PGD optimization steps per position, based on our hybrid heuristic–PGD strategy.

\vspace{-0.5em}
\subsubsection{Road Straightening Attack}
\textbf{Mapping Results.}
Table~\ref{tab:exp_rsa_map} presents AP results under four settings: clean conditions, random sampling baseline, PSO baseline, and our proposed Road Straightening Attack (RSA).
All attack types significantly degrade map quality compared to the clean setting. Road boundaries and lane dividers are the most affected elements; for example, $AP_{boundary}$ drops from 48.9\% to 39.0\% under adversarial patch attack, while $AP_{divider}$ drops from 54.2\% to 44.4\% under camera blinding. Although RSA primarily targets road boundaries, structural changes such as converting a turn into a straight road also affect nearby map elements, demonstrating broader map disruption.
Compared to the random sampling and PSO baselines, RSA consistently causes greater AP degradation, with up to 5.8\% more reduction in $AP_{boundary}$ and 4.5\% more in $AP_{divider}$. The only minor exception is AP on pedestrian crossing, which is acceptable since they are not the focus and are rare in the dataset. Overall, these results confirm that RSA effectively leverages model vulnerabilities and outperforms baselines under the same query budget. 
Notably, adversarial patches more strongly impact road boundaries, the main target of RSA, while camera blinding causes broader degradation across map element classes, leading to lower overall mAP. This highlights the broader disruption from camera blinding versus the targeted nature of adversarial patches.

\textbf{Planning Impact.} Table~\ref{tab:exp_rsa_plan} shows the effect of RSA on planning via Unreachable Goal Rate (UGR). Even in clean conditions, UGR is 27\%, revealing the model's inherent vulnerability with asymmetric scenes. Under our attack, UGR rises to 44\% for both flashlight and patch attacks—an absolute increase of 17\%, meaning nearly half the scenes become partially unreachable due to incorrect map predictions. 
Our method increases UGR by up to 13\% over the random sampling baseline and 10\% over the PSO baseline, demonstrating its effectiveness in blocking planned routes by exploiting symmetry bias. While PSO uses heuristic search to converge more efficiently than random sampling under the same query budget, our method causes even greater disruption.
As shown in Fig.~\ref{fig:exp_rsa_example}, both attack vectors mislead the model into predicting straight boundaries (green) instead of turns (black), leading to failed planning beyond the turn.

\subsubsection{Early-Turn Attack}
\textbf{Mapping Results.} Table~\ref{tab:exp_eta_map} shows that the Early Turn Attack leads to moderate degradation in map accuracy, with similar effects from camera blinding and adversarial patch attacks. Specifically, $AP_{boundary}$ drops from 48.9\% to 44.2\%, and $AP_{divider}$ from 54.2\% to 51.1\%. This smaller drop is expected, as ETA aims to subtly shift boundaries earlier rather than drastically alter road structure. Notably, $AP_{ped}$ drops more significantly, likely due to distortions in pedestrian crossings near turns. Overall, ETA outperforms the baselines by inducing a larger drop in $mAP$.

\textbf{Planning Impact.} As shown in Table~\ref{tab:exp_eta_plan}, ETA causes a sharp increase in Unsafe Planned Trajectory Rate (UPTR): from 10\% (clean) to 27\% (+17\%) with camera blinding and 21\% (+11\%) with adversarial patch. 
These exceed the random sampling baseline, which raise UPTR by only 12\% and 1\%, and also outperform the PSO baseline, which yields increases of 11\% and 4\%, respectively. This highlights that even small AP drops (1–3\%) can lead to significant planning errors and safety risks. Moreover, a larger AP drop doesn't necessarily translate to greater planning impact—for example, although the PSO baseline with blinding yields lower $AP_{boundary}$, it causes a smaller UPTR increase than ETA.
Compared to Road Straightening, ETA is more challenging to execute. It depends on reference boundaries after turns, which are farther from the vehicle and provide weaker visual cues than the closer opposite-side references used in RSA.
Fig.~\ref{fig:exp_eta_example} visualizes ETA’s impact on planning, where both attack vectors induce subtle but critical early boundary shifts, resulting in boundary violations. Fig.~\ref{fig:exp_eta_patch} shows an example where the model incorrectly predicts symmetric turning boundaries, indicating that symmetry bias extends beyond straight-road predictions.

\vspace{-0.5em}
\subsection{Attack Transferability}
We evaluate the transferability of our attack by targeting a different model. In this experiment, the victim AV uses VectorMapNet for online map construction, while the attack configurations are computed using MapTR on the same set of 100 asymmetric scenes.
As shown in Appendix~\ref{sec:attack_transfer_appendix}, the attack remains effective with reduced impact: up to 4.3\% drop in boundary AP, 3.6\% in lane divider AP, and 3.5\% in mAP. It also increases the Unreachable Goal Rate by up to 15\%, while Unsafe Planned Trajectory Rate shows no significant change. These results demonstrate the attack’s cross-model transferability, particularly for road straightening, which is easier to trigger.

\vspace{-0.5em}
\subsection{Attack Generalizability}
\textbf{Attack Other State-of-the-art Models.} We apply the Road Straightening and Early Turn Attacks to two state-of-the-art industry-proposed models: GeMap~\cite{zhang2023online}, which uses geometric representations of map elements, and MapQR~\cite{liu2024leveraging}, which enhances map query capabilities. As shown in Table~\ref{tab:exp_rsa_advanced} and Table~\ref{tab:exp_eta_advanced} in the Appendix, both attacks significantly degrade performance, causing notable drops in map AP and reaching up to 39\% Unreachable Goal Rate and 31\% Unsafe Planned Trajectory Rate. These results suggest that the vulnerability stems from multiple root causes and cannot be effectively addressed by improved model design alone.

\textbf{Attack LiDAR-camera Fusion-based Model.}\label{sec:attack_lidar_camera_fusion} While most online map construction models are vision-only, some fuse multi-view camera images with LiDAR point clouds to leverage LiDAR's precise 3D information. Our attack employs flashlight and adversarial patch vectors that perturb only camera inputs, leaving LiDAR unaffected. However, as shown in the \textit{C \& L} rows of Table~\ref{tab:exp_rsa_advanced} and Table~\ref{tab:exp_eta_advanced} in the Appendix, despite reduced impact compared to camera-only models, the attack still achieves up to 27\% Unreachable Goal Rate and 18\% Unsafe Planned Trajectory Rate on GeMap with LiDAR-camera fusion. This indicates that camera inputs—susceptible to our attacks—remain critical for map element recognition, and that adding additional sensing modalities alone does not defend against our attacks. Further discussion on attacking fusion-based models is presented in Section~\ref{sec:discussion_attack_fusion}.

\textbf{Attack End-to-End AD Model.} End-to-end (E2E) autonomous driving models are increasingly popular, often incorporating online map construction modules with architectures similar to models like MapTR. We apply the Road Straightening and Early Turn attacks to VAD~\cite{jiang2023vad}, a widely used E2E model, across 100 selected asymmetric scenes.
To evaluate the model’s online map construction, we use standard map metrics on its auxiliary map output. For planning, unlike the rule-based Hyper A*-planner, E2E models predict the ego vehicle's trajectory over the next 3 seconds. Accordingly, we use the average L2 distance, a common metric for E2E planning evaluation.
As shown in Table~\ref{tab:attack_e2e} in the Appendix, mapping performance drops sharply, with mAP falling from $50.8\%$ to $19.4\%$. The average L2 distance increases from $0.77m$ to $3.71m$, indicating substantial trajectory deviation.
These results show that our attacks not only compromise specialized online map construction models but also significantly impair E2E autonomous driving systems.

\section{Experiments on Real-World}\label{sec:exp_realworld}
\begin{figure}[!t]
    \centering
    \hfill
    \subfigure[Testbed AV.]{
        \label{fig:exp_testbed_av}
        \includegraphics[width=0.43\columnwidth]{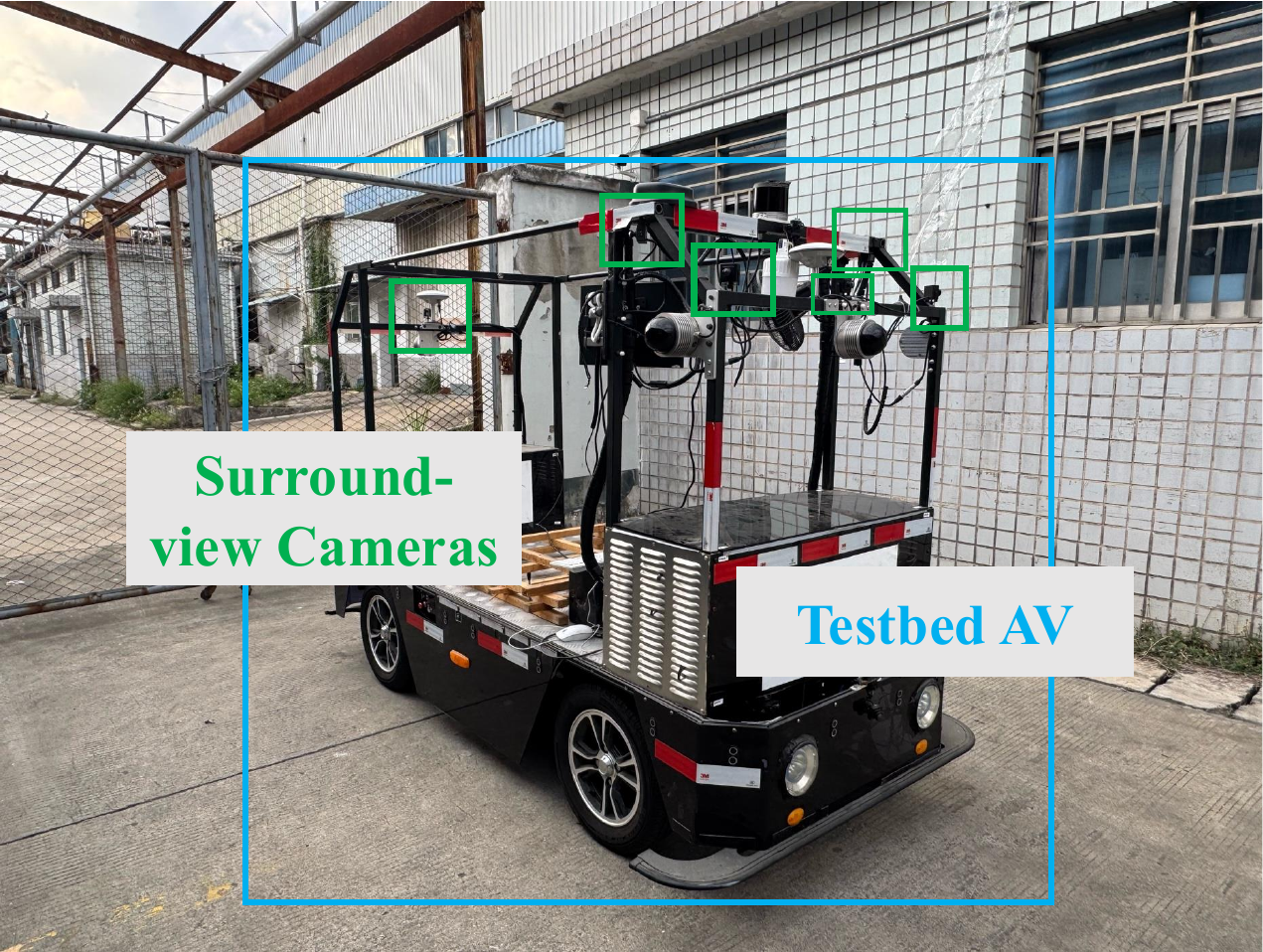}
    }\hfill
    \subfigure[Attack Vectors.]{
        \label{fig:exp_attack_vectors}
        \includegraphics[width=0.43\columnwidth]{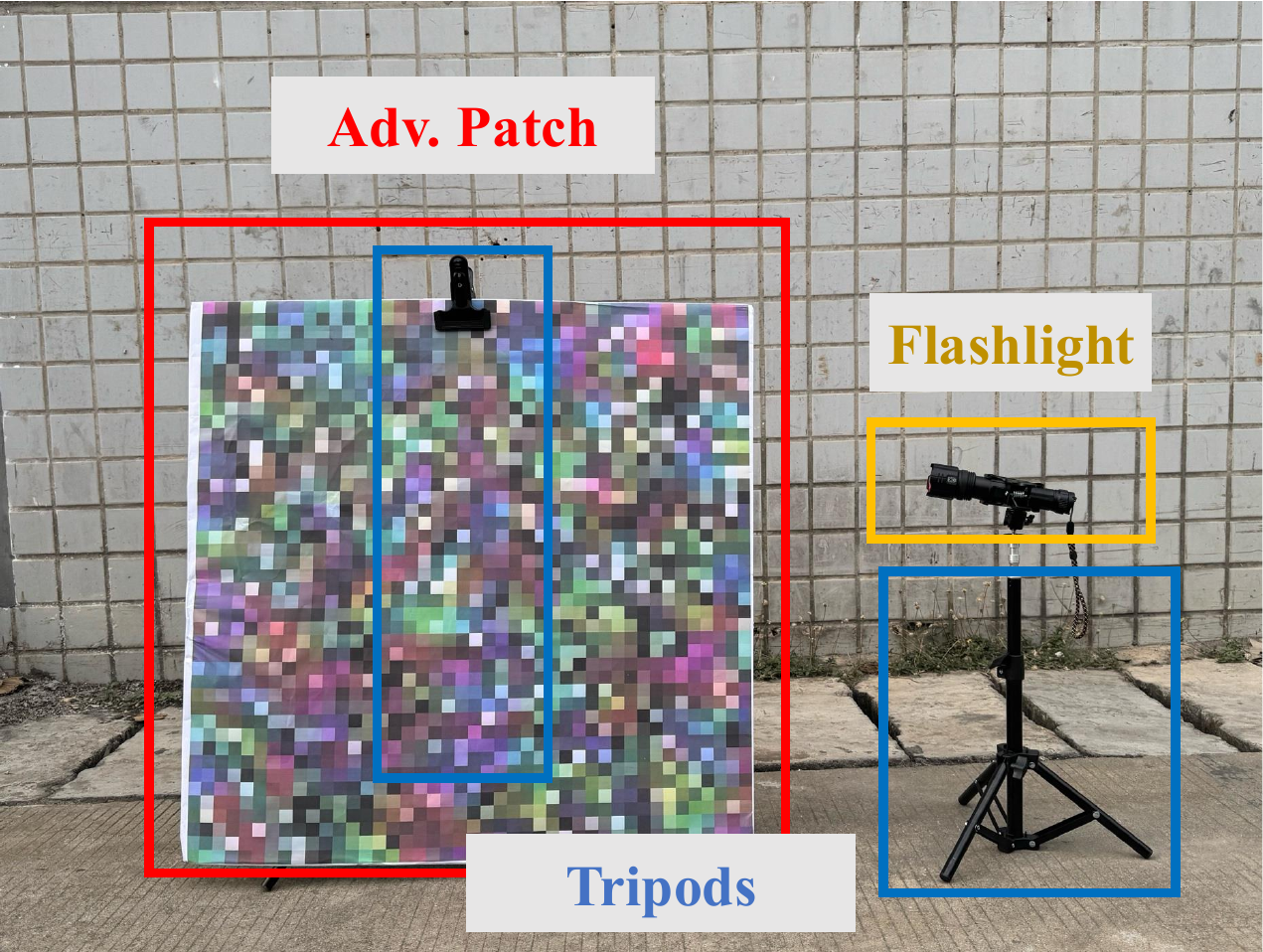}
    }\hfill
    \vspace{-1em}
    \caption{Real-world Experiment Setup.}
    \label{fig:camera_in_car}
    \vspace{-1em}
\end{figure}

\subsection{Experiment Setting}
\textbf{Testbed AV.} 
As shown in Fig.~\ref{fig:exp_testbed_av}, our testbed AV is equipped with six surround-view cameras mounted at the front, front-left, front-right, back, back-left, and back-right positions. Each camera has a wide-angle lens with a 180° horizontal and 60° vertical field of view, enabling full 360° horizontal coverage and sufficient vertical visibility.
Cameras capture images at 800×600 resolution and 10 Hz, with hardware synchronization to ensure temporal alignment. All cameras are calibrated to obtain intrinsic and extrinsic parameters, ensuring spatial consistency.
The testbed AV runs the same MapTR model used in our dataset experiments for online map construction.

\begin{figure*}[!t]
    \centering
    \hfill
    \subfigure[Camera views under road straightening attack. The front and front-right views are obscured by a flashlight.]{
        \label{fig:real_exp_straghtening_cam}
        \includegraphics[width=0.3\textwidth, trim=0 70 0 40, clip]{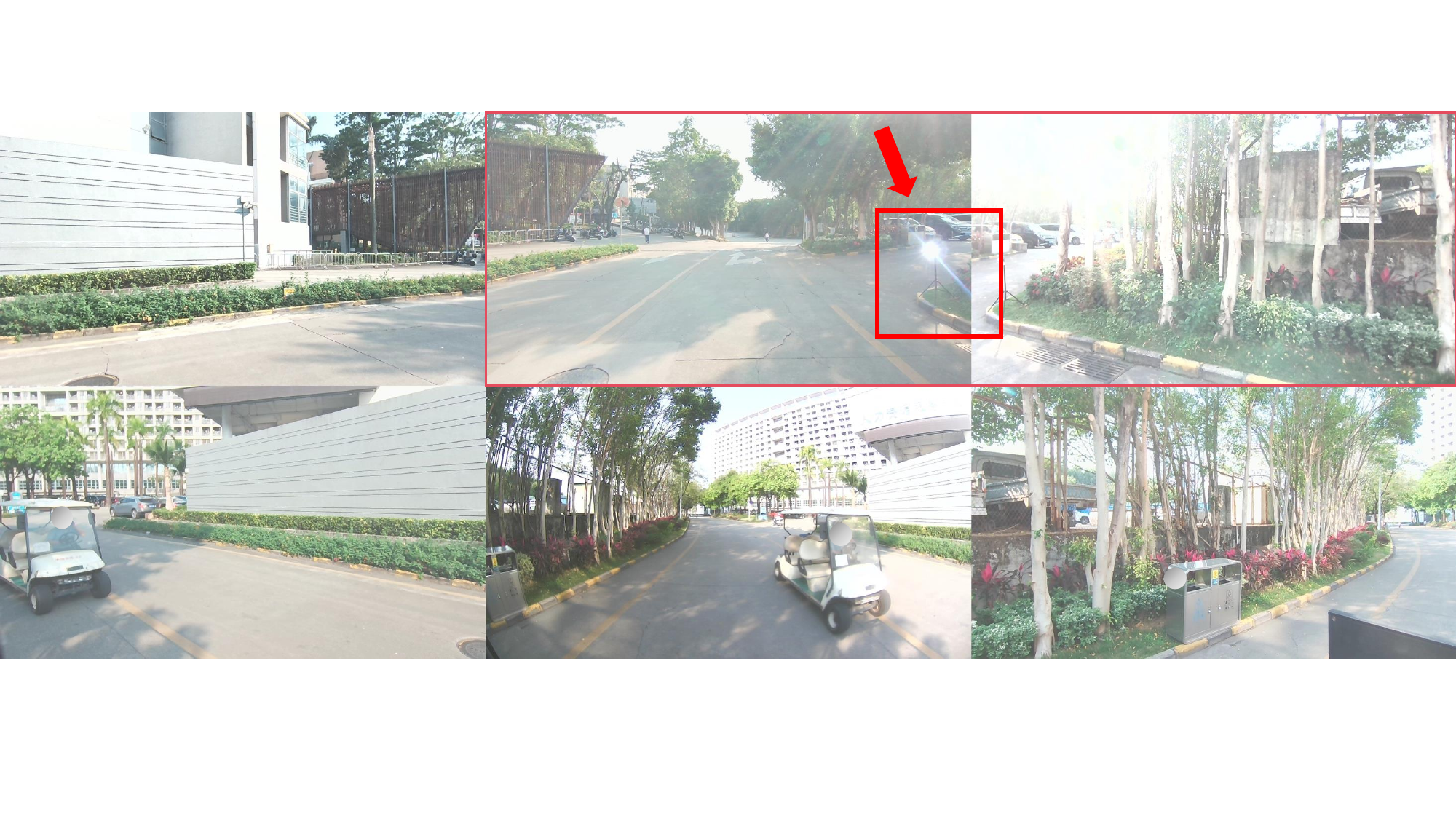}
    }\hfill
    \subfigure[Map result in clean condition. Success ratio: 1/8.]{
        \label{fig:real_exp_straghtening_clean}
        \includegraphics[width=0.2\textwidth]{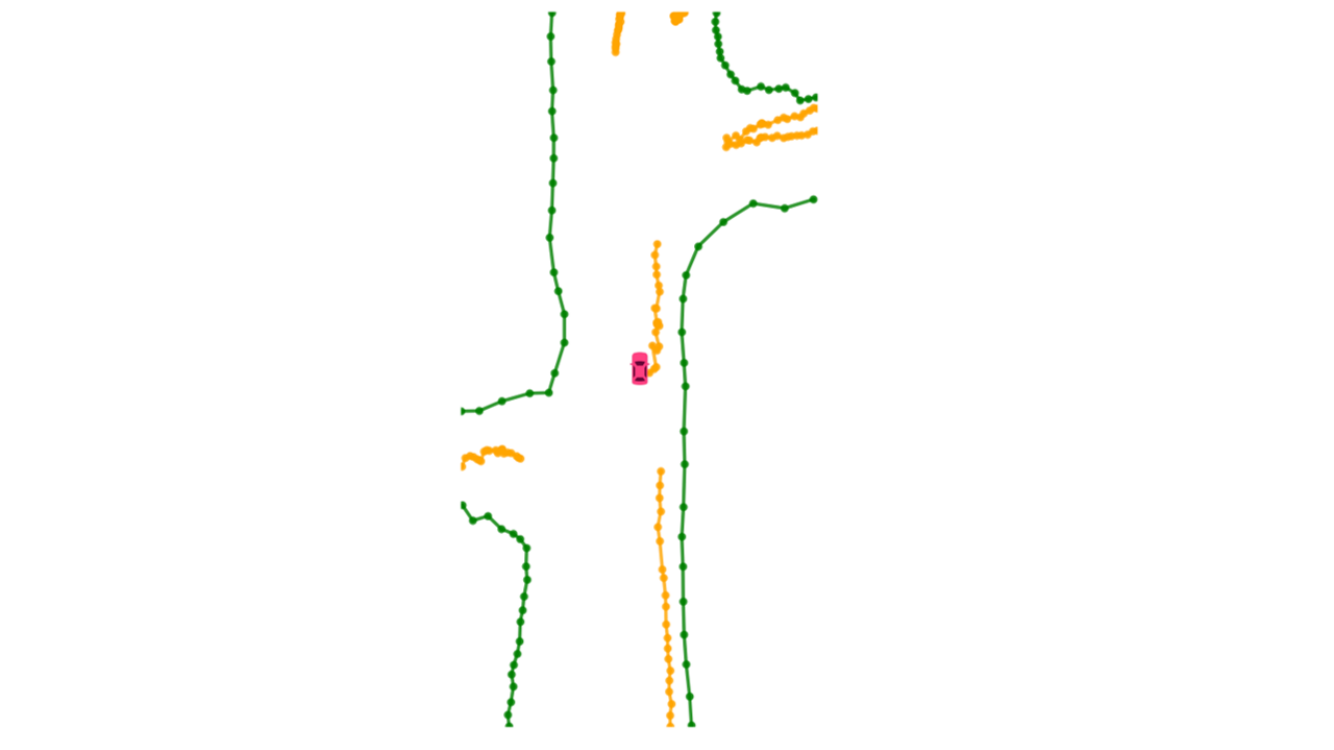}
    }\hfill
    \subfigure[Map under road straightening attack via blinding. Success ratio: 8/8.]{
        \label{fig:real_exp_straghtening_blind}
        \includegraphics[width=0.2\textwidth]{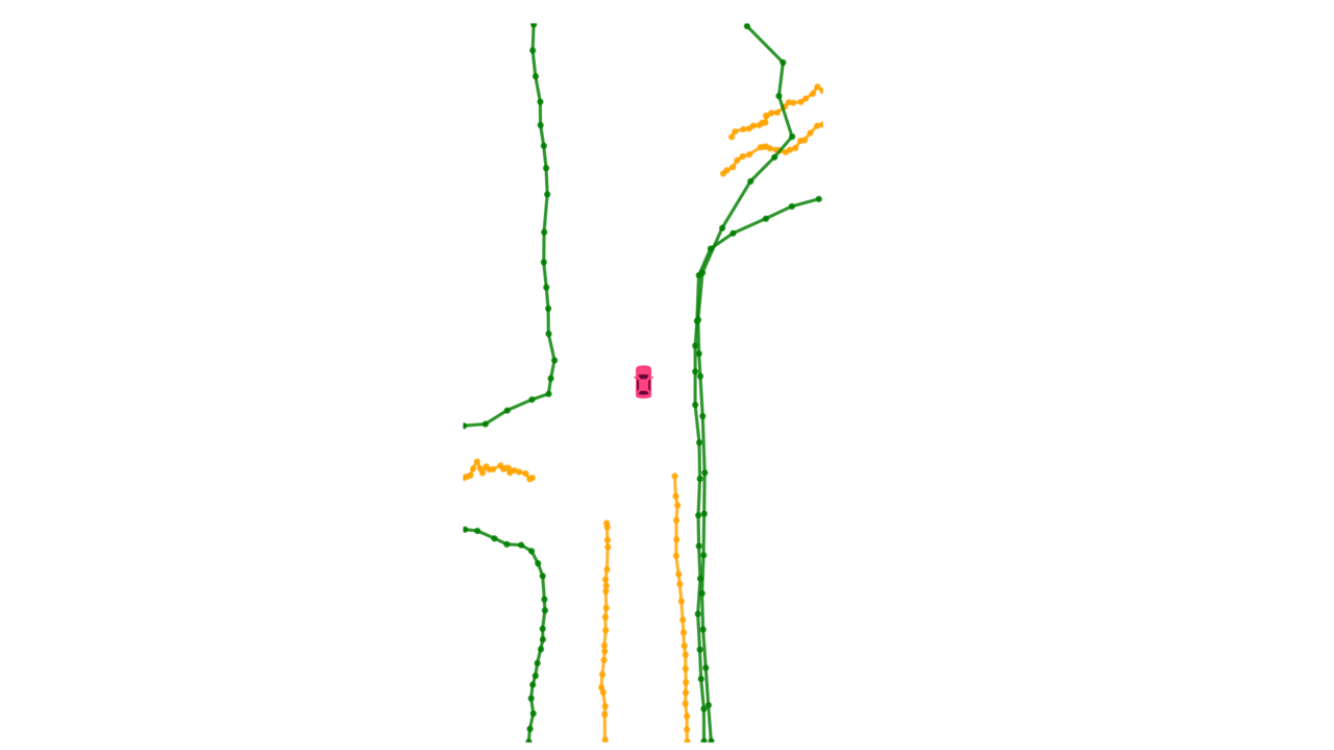}
    }\hfill
    \subfigure[Map under road straightening attack via adv. patch. Success ratio: 7/8.]{
        \label{fig:real_exp_straghtening_patch}
        \includegraphics[width=0.2\textwidth]{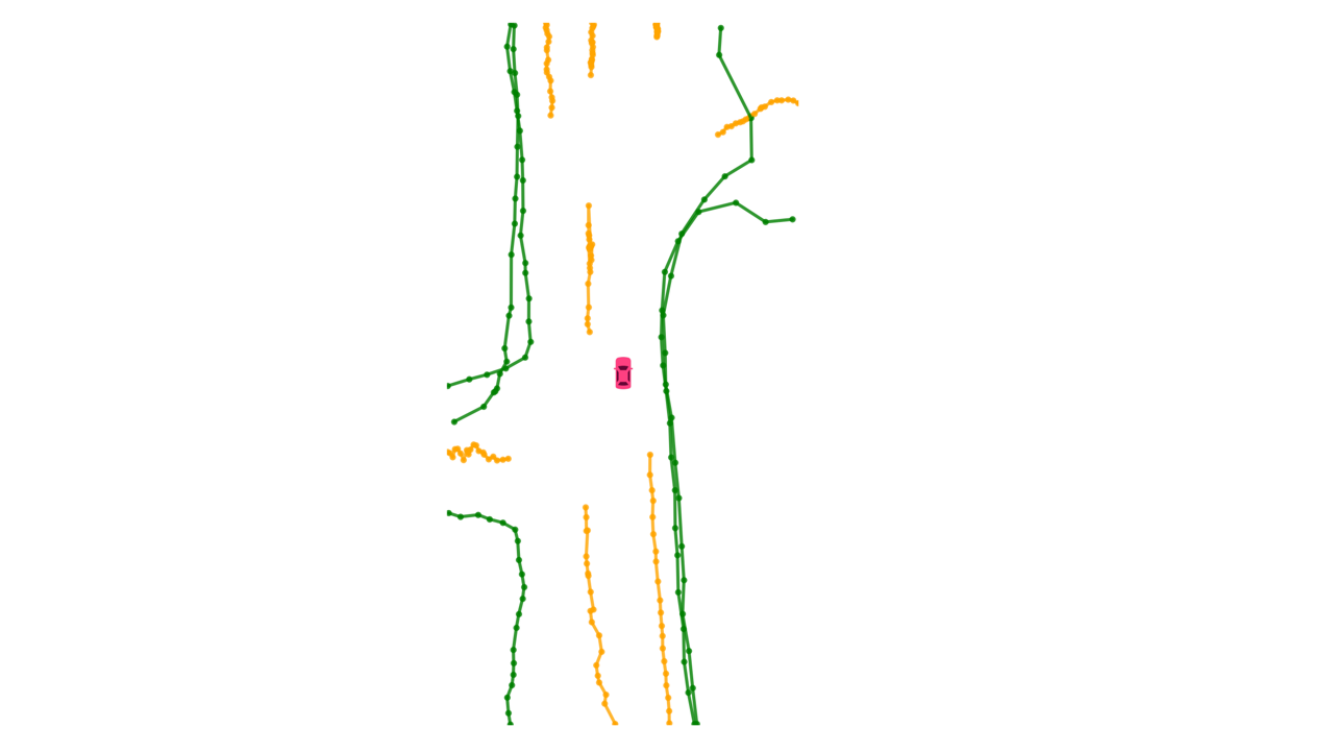}
    }\hfill\\
    \hfill
    \subfigure[Camera views under early turn attack. The patch is visible in the front camera view.]{
        \label{fig:real_exp_earlyturn_cam}
        \includegraphics[width=0.3\textwidth, trim=0 70 0 40, clip]{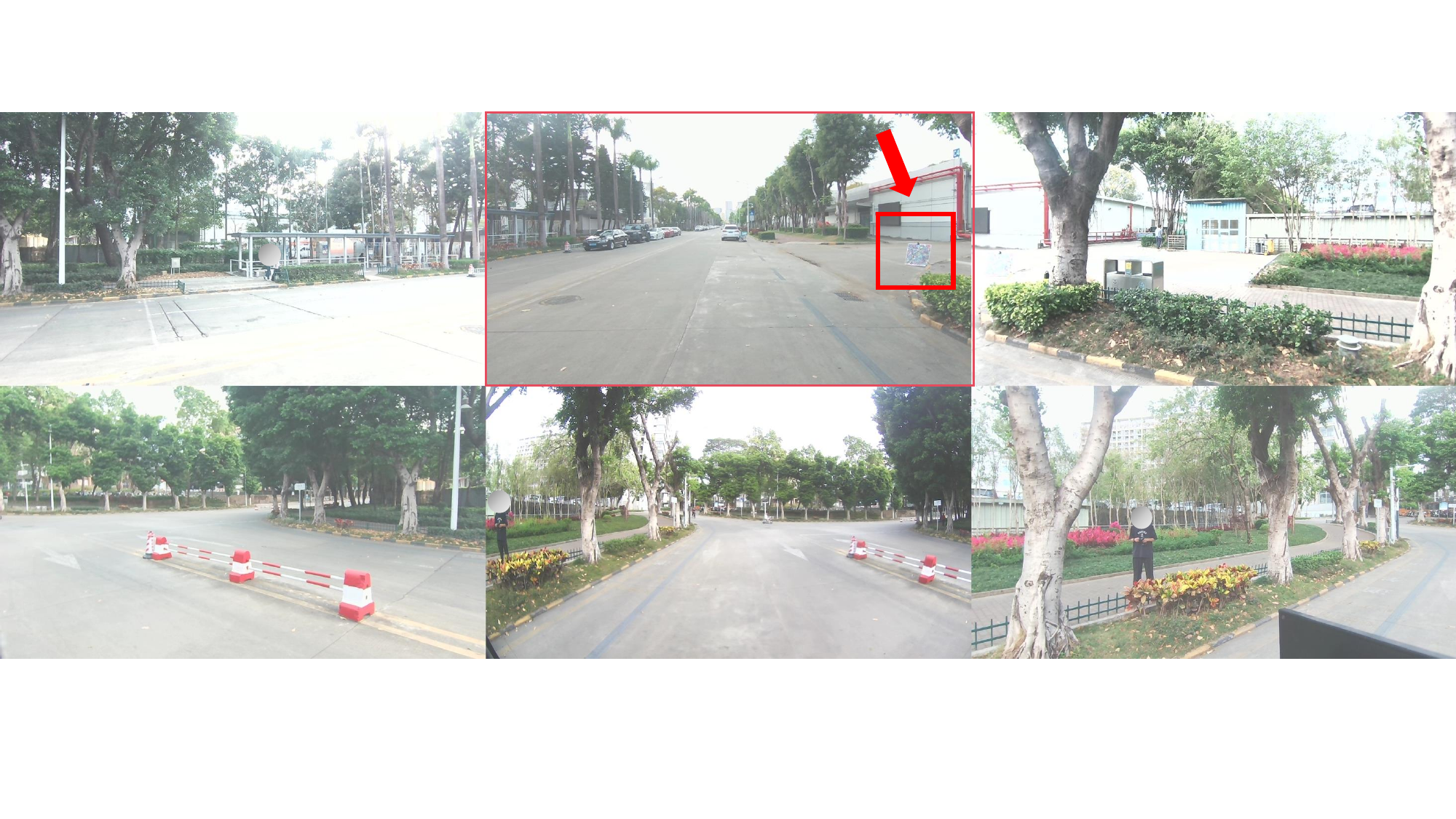}
    }\hfill
    \subfigure[Map result in clean condition. Success ratio: 0/8.]{
        \label{fig:real_exp_earlyturn_clean}
        \includegraphics[width=0.2\textwidth]{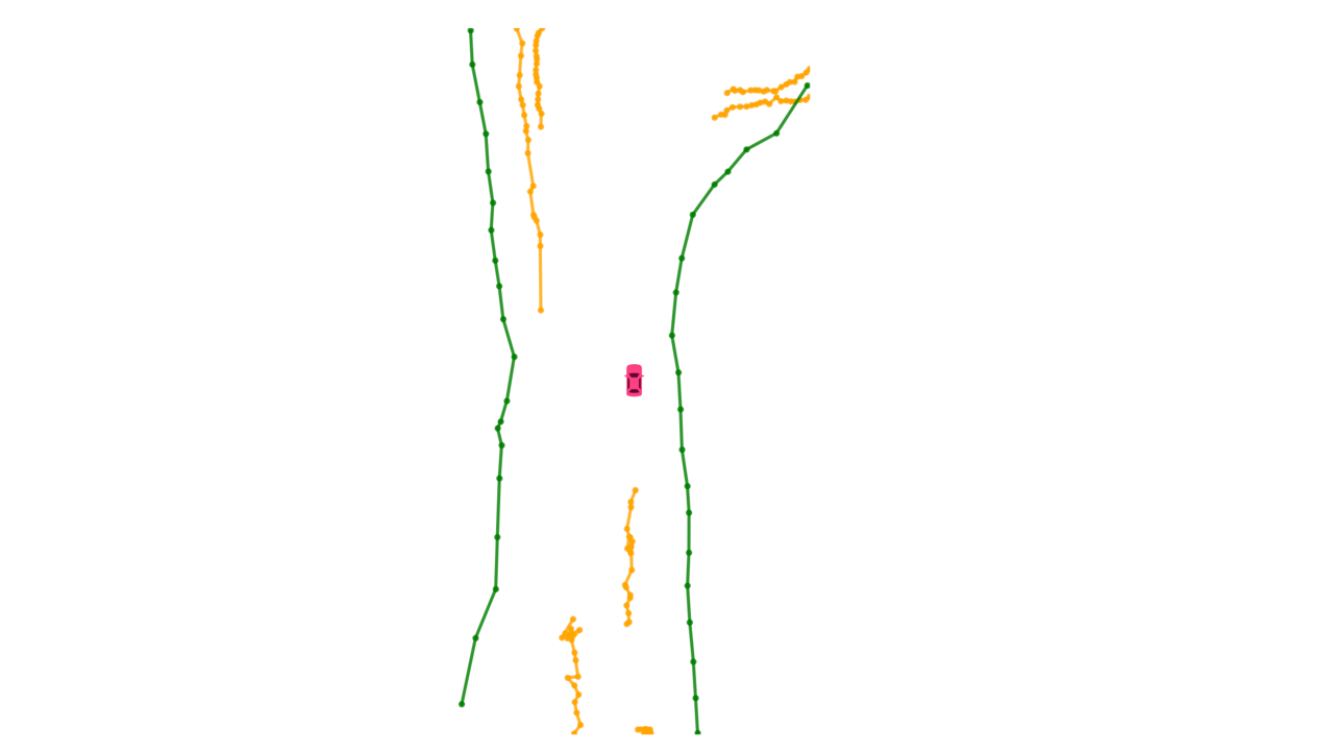}
    }\hfill
    \subfigure[Map under early turn attack via blinding. Success ratio: 3/8.]{
        \label{fig:real_exp_earlyturn_blind}
        \includegraphics[width=0.2\textwidth]{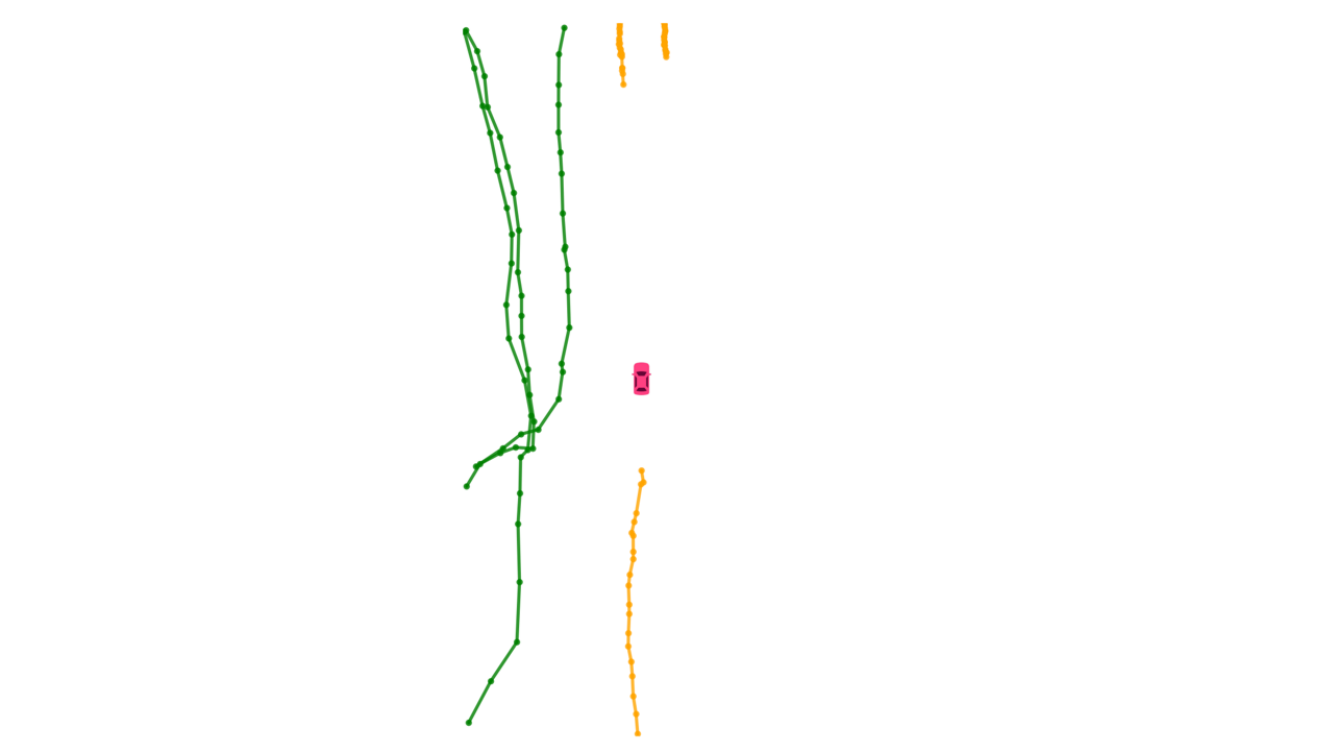}
    }\hfill
    \subfigure[Map under early turn attack via adv. patch. Success ratio: 2/8.]{
        \label{fig:real_exp_earlyturn_patch}
        \includegraphics[width=0.2\textwidth]{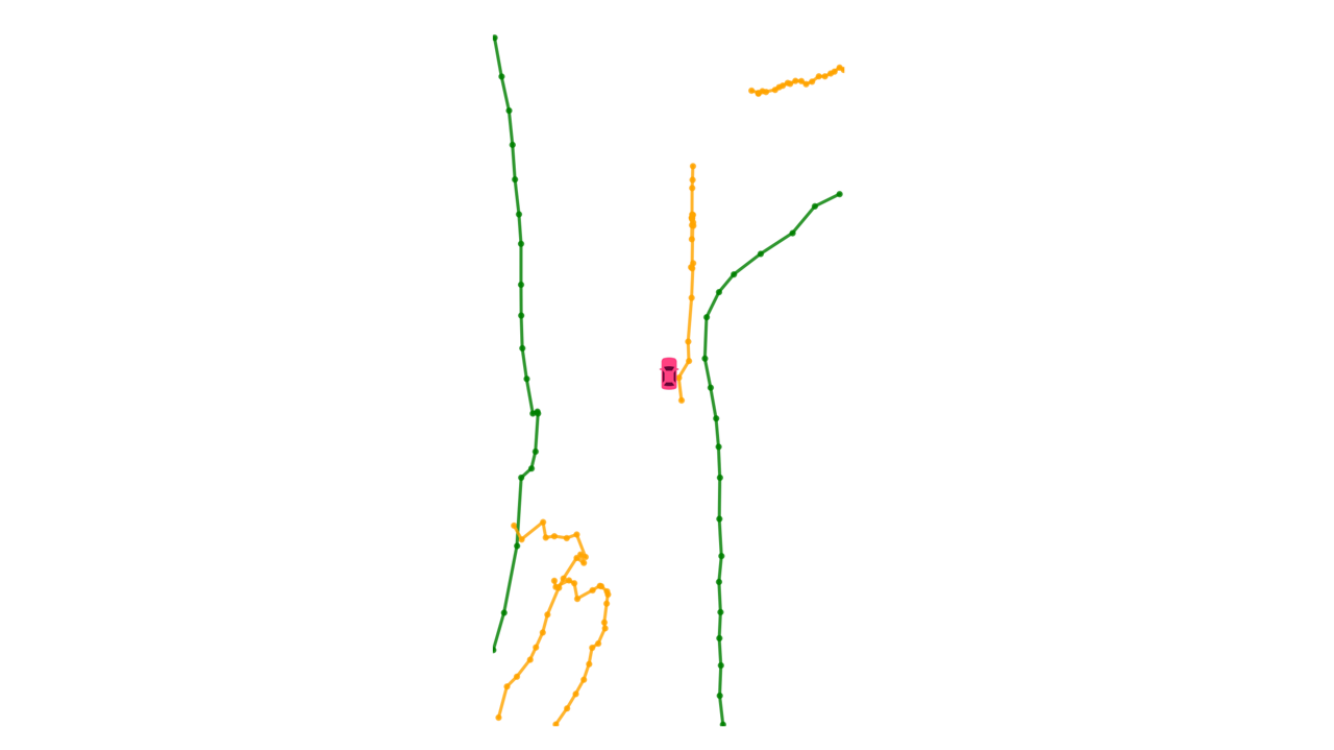}
    }
    \hfill
    \vspace{-0.5em}
    \caption{Two real-world attack scenarios. Top row: Road Straightening Attacks; Bottom row: Early Turn Attacks.}
    \label{fig:real_world_attack}
\end{figure*}

\begin{figure}[!t]
    \centering
    \hfill
    \subfigure[Road merge: clean vs. Road Straightening Attack using an adversarial patch.]{
        \label{fig:exp_road_merge}
        \includegraphics[width=0.45\columnwidth]{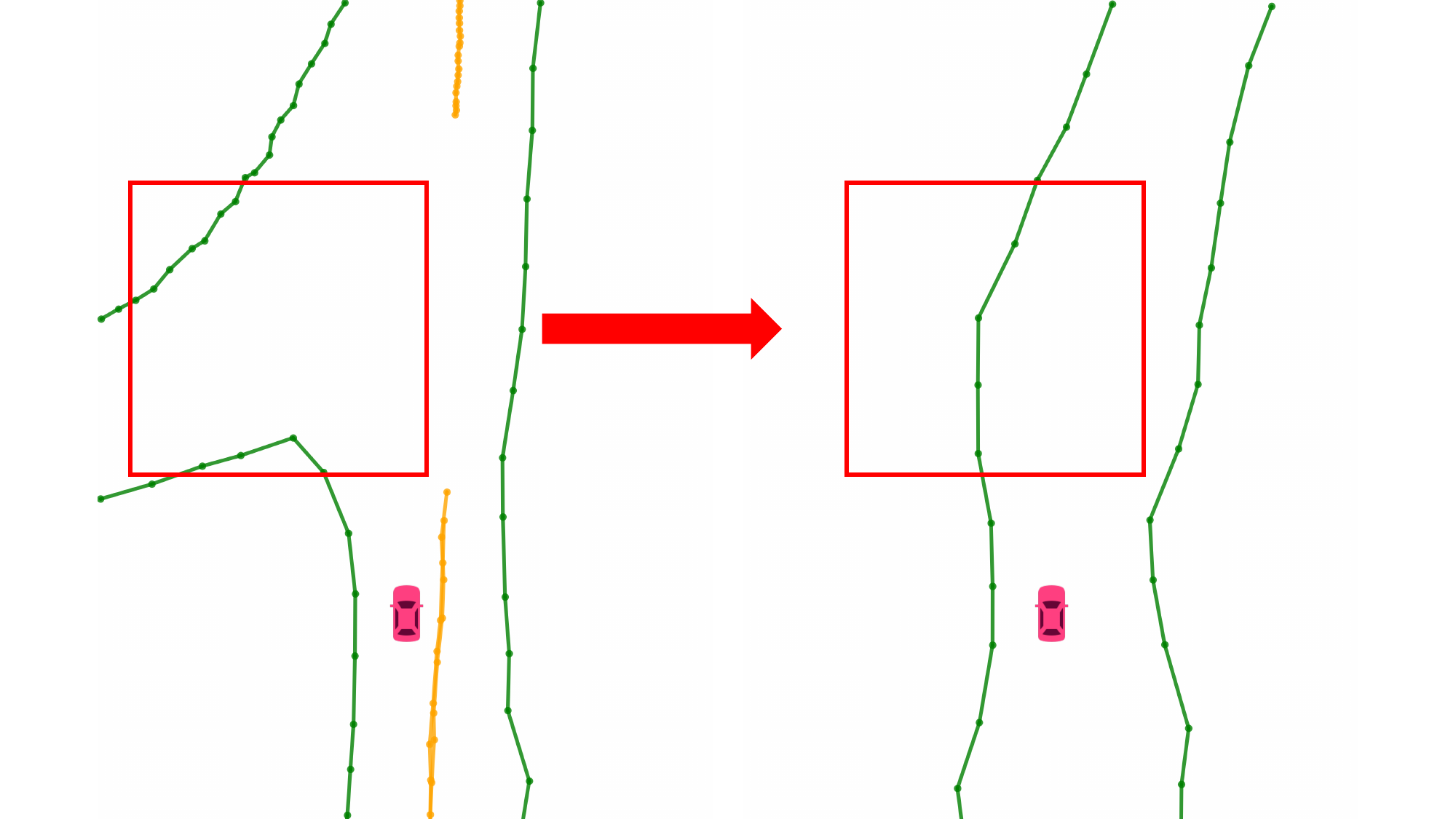}
    }\hfill
    \subfigure[Road split: clean vs. Early Turn Attack using camera blinding.]{
        \label{fig:exp_road_split}
        \includegraphics[width=0.45\columnwidth]{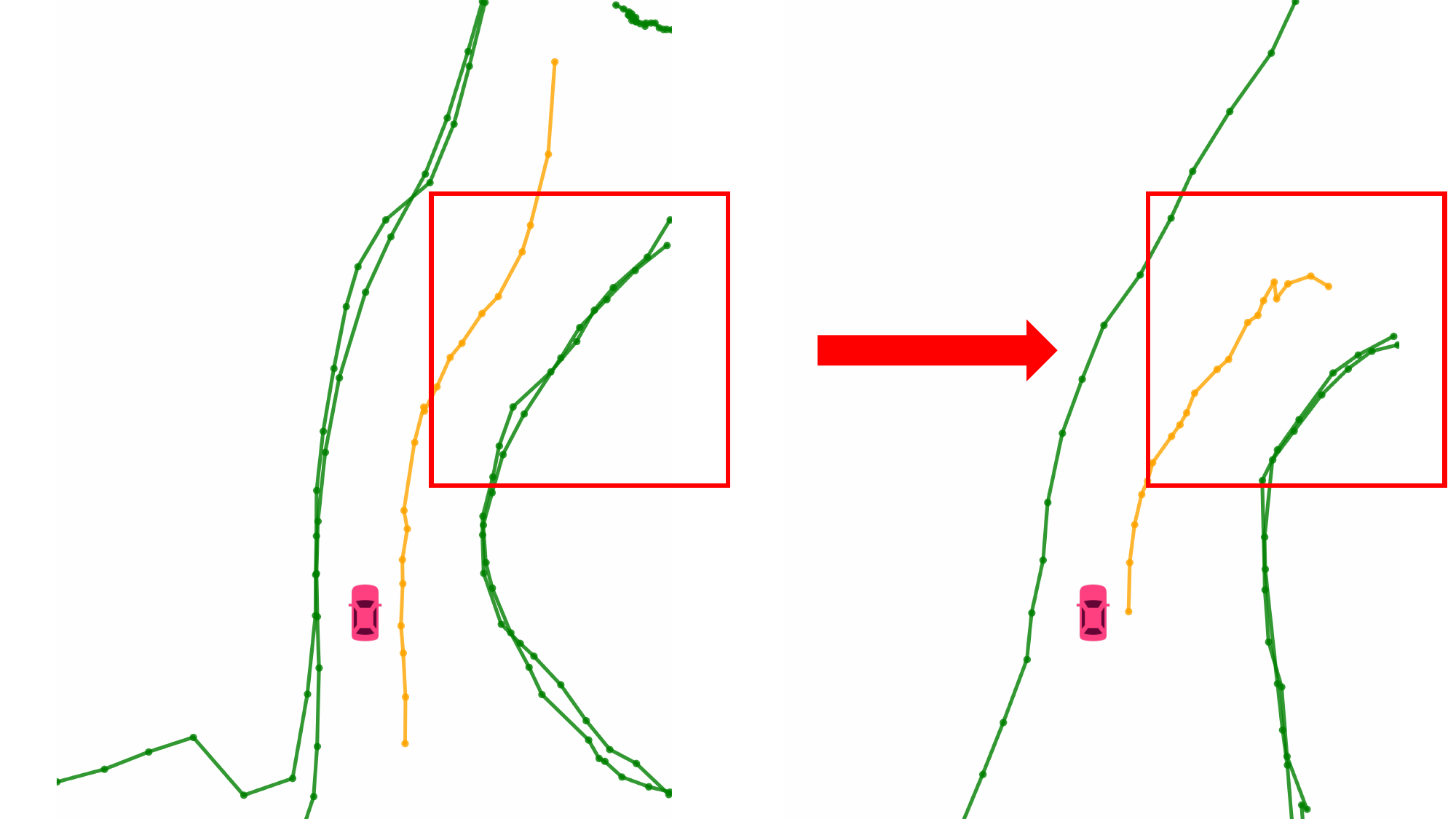}
    }\hfill
    \vspace{-0.8em}
    \caption{Example map results under successful attacks on asymmetric road types, including road merges and splits.}
    \label{fig:road_types}
    \vspace{-1em}
\end{figure}


\textbf{Flashlight and Adversarial Patch.} 
As shown in Fig.~\ref{fig:exp_attack_vectors}, we use a custom attack toolkit consisting of two telescopic tripods, a flashlight, and an adversarial patch board.
The flashlight is a commercially available model with 3,000 lumens peak brightness and adjustable focus. During attacks, we use the highest brightness setting and fine-tune the focus to maximize intensity and lens flare effect.
The adversarial patch is printed on a $1\mathrm{m} \times 1\mathrm{m}$ board using precomputed patterns. Both the flashlight and patch are mounted on height-adjustable tripods (55–210 cm) for flexible placement and orientation based on the optimized attack configuration.
This setup is low-cost, roadside-deployable, and practical for real-world attacks against online map construction models.

\textbf{Attack Planning and Deployment.}
Since no SD/HD map is available for the test site, we first construct the map ourselves. The testbed AV, acting as the attacker's data collection vehicle, drives through the site to generate online constructed maps. 
Using our asymmetric scene identification method, we identify and select eight representative asymmetric scenes, covering road types such as forks, splits, and merges.
For each scene, we select one target frame where the AV approaches the asymmetry anchor. Using the collected data, our framework generates four attack configurations per scene: flashlight positions for Road Straightening and Early Turn attacks, and adversarial patch positions (including facing angle) and patterns for both attack types.
In the deployment phase, we mount the flashlight on a tripod at the selected position for the camera blinding attack. For the adversarial patch, we print the pattern, attach it to a board, and mount it on a tripod at the precomputed position and angle. The testbed AV then replays each scene as the victim, collecting surround-view images at the same location under clean and attack conditions. These images are fed into the online map construction model for evaluation.

\vspace{-0.5em}
\subsection{Attack Effectiveness}
\subsubsection{Road Straightening Attack}
In clean conditions, the model correctly predicts asymmetric turns in 7 out of 8 scenes. As shown in Fig.~\ref{fig:real_exp_straghtening_cam}, the right boundary curves into a parking lot while the left remains straight—a typical asymmetric fork layout. The model captures this correctly in clean condition (Fig.~\ref{fig:real_exp_straghtening_clean}). However, in one case (see Fig.~\ref{fig:real_world_clean_fail} in Appendix), it incorrectly predicts a straight boundary without interference, revealing an inherent symmetry bias.
When applying the Road Straightening Attack using a flashlight, we achieve a 8/8 success rate. In Fig.~\ref{fig:real_exp_straghtening_blind}, the flashlight, partially obscures the right turn in the front and front-right camera views. The model mispredicts the curved road as straight, preventing the AV from entering the parking lot. The adversarial patch achieves 7/8 successful attacks, producing similar road straightening effects. An example is shown in Fig.~\ref{fig:real_exp_straghtening_patch}. 
Beyond forks, the Road Straightening Attack is also effective on other asymmetric road types. For example, in a road merge scene (Fig.~\ref{fig:exp_road_merge}), the adversarial patch misleads the model into ignoring a merging lane, potentially leaving the victim AV unaware of fast-approaching traffic—posing a safety risk.
Overall, both attack vectors achieve high success rates, confirming that the vulnerability persists and is easily triggered in real-world settings. Most effective attack positions are located near the start of the turn, around asymmetry anchors, highlighting their importance in attack planning.

\subsubsection{Early Turn Attack}
In clean conditions, all 8 scenes are correctly predicted with no early turns. An example is shown in Fig.~\ref{fig:real_exp_earlyturn_cam}, where the right turn appears clearly in the camera view and is accurately predicted in map result (Fig.~\ref{fig:real_exp_earlyturn_clean}).
Using the flashlight, 3 out of 8 scenes result in successful early turn attacks. In one case (Fig.~\ref{fig:real_exp_earlyturn_blind}), the right boundary is entirely missing—considered a valid early turn outcome—raising the risk of curb collisions.
The adversarial patch attack achieves 2 successes. In Fig.~\ref{fig:real_exp_earlyturn_patch}, a patch placed after the turn causes the model to predict the boundary too early, creating a similar hazard.
Fig.~\ref{fig:exp_road_split} illustrates an Early Turn Attack on a road split. The camera blinding causes the model to predict an early right-turn boundary, which could mislead the victim AV into colliding with the actual turn boundary when attempting a lane change.
While effective across various asymmetric road types, early turn attacks are generally harder to trigger than road straightening. This may be due to varying sunlight during deployment, which affects patch visibility. We also find that effective attack positions for ETA are typically just before or after the turn, but still around the asymmetry anchors.

\subsection{Attack Robustness}
\textbf{Attack under Different Weather Conditions.} 
We evaluate the effectiveness of the Road Straightening Attack (RSA) and Early Turn Attack (ETA) in two cloudy and two nighttime scenarios.
Representative scenarios are shown in Fig.~\ref{fig:weather} in Appendix.
In cloudy conditions, RSA succeeds in 2 out of 2 scenes for both camera blinding and adversarial patch attacks. ETA succeeds in 1 out of 2 scenes using camera blinding, but fails in both scenes when using adversarial patches.
At night, RSA with camera blinding remains effective, while all other attacks fail.
Camera blinding is significantly more effective in low-light conditions, such as nighttime in Fig.~\ref{fig:exp_night}, due to enhanced blur and glare from the flashlight, which more effectively obscures critical regions like asymmetry anchors. In contrast, adversarial patches become nearly imperceptible at night, significantly reducing their effectiveness under such conditions.

\textbf{Attack in Traffic.} 
In real-world scenarios, attacks may be affected by other road agents such as vehicles or pedestrians, especially in asymmetric scenes like forks where traffic tends to concentrate. To evaluate the attack robustness in such conditions, we perform the road straightening attack using camera blinding in two traffic scenarios: one involving pedestrians crossing the road, and another with a passing vehicle. 
As shown in Fig.~\ref{fig:traffic_ped}, the attack remains effective as long as pedestrians do not block the victim AV’s view of the attack vectors. 
Interestingly, Fig.~\ref{fig:traffic_vehicle} shows that even temporary occlusion of asymmetry anchors by passing vehicles alone can trigger the erroneous symmetric predictions. This highlights that obstructing critical asymmetric regions is key to triggering the vulnerability—consistent with the mechanisms of both camera blinding and adversarial patch attacks.

\vspace{-0.3em}
\section{Root Cause Analysis and Defenses}\label{sec:root_causes}

Our results demonstrate that symmetry bias constitutes a vulnerability exploitable by the attacks proposed in Section~\ref{sec:method}.
In this section, we analyze three root causes of this vulnerability and introduce asymmetric data fine-tuning as a potential defense to mitigate the threat.

\textbf{Cause 1: Training Data Imbalance.}
Using the method from Section~\ref{sec:method_identify_asym}, we identify 407 asymmetric scenes out of 2,095 in the nuScenes validation set and 2,471 out of 10,305 in the training set, both around 20\% of the total data. This reveals a data imbalance, with symmetric scenes dominating the dataset (and real-world driving), which likely contributes to the model’s symmetry bias.
To examine whether data imbalance reinforces symmetry bias, we measure model uncertainty using Monte Carlo Dropout~\cite{gal2016dropout} on 100 symmetric and 100 asymmetric scenes. 
We observe that asymmetric scenes misclassified as symmetric show notably lower model uncertainty (0.178), suggesting the model is overconfident in its incorrect symmetric predictions. In contrast, it shows higher uncertainty (0.464) when generating asymmetric predictions, indicating unfamiliarity with such predictions.

\textbf{Cause 2: Network Design.} 
As discussed earlier, online map construction models typically use a BEV encoder–map decoder architecture. In the decoder, such as in MapTR (see Appendix Fig.~\ref{fig:root_cause_network_design}), the query for a diverging boundary interacts with both instance-level queries (e.g., reference boundaries) and point-level queries (e.g., points along the same boundary). This design allows straight reference boundaries or pre-turn points on the diverging boundary to inject misleading contextual cues, often causing the model to generate symmetric predictions that mirror those geometries.

\textbf{Cause 3: Map Element Representation.} 
Online map construction models represent vectorized map elements as polylines or polygons, typically using 20 unconstrained points. While this representation is flexible to captures complex structures like S-curves, it also makes the geometry easy to manipulate, e.g., turning a curve into a straight or irregular line.

\textbf{Defense: Asymmetric Data Fine-tuning.}
Improving network design or output format is beyond the scope of this work. 
To address the imbalance in training data, we fine-tune the pre-trained MapTR model on 2,471 identified asymmetric training frames for an additional 10 epochs.
On the 100 asymmetric evaluation scenes (see Appendix~\ref{sec:defense_appendix}), the fine-tuned model improves road boundary AP by 5.2\% under clean conditions and achieves up to 11\% reduction in Goal Unreachable Rate and 3\% reduction in Unsafe Planned Trajectory Rate under attack, demonstrating partial effectiveness as a defense.

\section{Discussion and Future Work}\label{sec:limitations}
\textbf{Continuous attack across frames.} 
A limitation of our attack is that its effect weakens once the victim AV moves past the flashlight or adversarial patch, as the visual interference fades. Future works could explore continuous attacks. For example, using pan-tilt units with flashlights to dynamically track and target the AV’s cameras in real time. Similarly, globally optimizing adversarial patterns across frames could help maintain their effectiveness throughout the vehicle’s trajectory.

\textbf{Stealthiness of Attack Vectors.}
To improve stealth, future attacks could use invisible lights, such as infrared lights instead of flashlight or adversarial patches, which can affect camera inputs without being visible to humans. Prior works~\cite{wang2021can, sato2024invisible, guo2024invisible} show that such attacks can disrupt vision-based tasks like traffic sign recognition and SLAM, suggesting potential effectiveness for online mapping systems.

\textbf{Attack Multi-sensor Fusion.}\label{sec:discussion_attack_fusion}
As discussed in Section~\ref{sec:attack_lidar_camera_fusion}, our attacks exhibit reduced impact on LiDAR-camera fusion-based online map construction models. However, our camera-only attack vectors can be extended to multi-sensor fusion systems.
For instance, camera blinding can be combined with LiDAR/Radar spoofing~\cite{jin2023pla, cao2019adversarial, vennam2023mmspoof} to inject false symmetric boundaries or remove diverging ones. Additionally, adversarial objects~\cite{cao2021invisible, zhu2024malicious} with crafted textures and shapes can be used to simultaneously introduce visual and point cloud perturbations, following the patch optimization procedure.

\textbf{Attack Commercial AD Systems.}
Commercial AD systems with online map construction module remain closed-source, but the asymmetry vulnerability broadly applies for two reasons: (1) the fundamental root causes—real-world data imbalance (dominant symmetric roads) and deep map element interaction designs—are difficult to eliminate, and (2) their planning modules rarely validate map inputs rigorously. Extending our attack to commercial systems may require simulating additional data processing steps used by these AD systems, such as sensor noise filtering and trajectory smoothing, to optimize for a more robust attack configuration.

\section{Conclusion}\label{sec:conclusion}
In this paper, we identify a model-level vulnerability in online map construction models: a bias toward predicting symmetric road structures. This bias can be exploited in asymmetric scenes, such as forks and merges, through physical interference, causing the model to mispredict them as symmetric (e.g., straight roads or intersections).
To exploit this vulnerability, we propose a two-stage attack framework that automatically detects vulnerable asymmetric scenarios and optimizes real-world attack configurations in both black-box and white-box settings. By deploying flashlights or adversarial patches based on the identified configuration at the roadside, our method misleads the victim AV into generating incorrect map predictions, leading to unsafe planning behaviors.
Evaluations on a public autonomous driving dataset and a real-world testbed AV show that our attacks significantly degrade map accuracy, render target routes unreachable, and increase collision risks.
We further analyze the root causes of this vulnerability and propose a defense to improve model robustness.


\section*{Ethical Consideration}\label{sec:ethical}
\textbf{Experimental safety measures:} 
The real-world experiments were safely conducted on closed roads in an industrial park under controlled conditions and with special permission. Three team members ensured safety: one operated the testbed AV, another monitored pedestrians and traffic and held a remote emergency braking controller for the testbed AV, ready to intervene at any time if a dangerous situation arose, and the third managed the attack vectors. During the flashlight attack, the team member responsible for positioning the flashlight also ensured that pedestrians were not exposed to strong light to protect their eye safety.

\textbf{Vulnerability disclosure:} 
We recognized that our research may impact autonomous driving companies and automakers. We have provided detailed information about the identified vulnerability and potential countermeasures to the maintainers of the affected models and the relevant companies. These details will also be published on our demo website.

\section*{Acknowledgments}
We thank the anonymous reviewers for their insightful feedback. 
We thank Jinghuai Deng, Jie Wang, Tianchi Ren (CityU of Hong Kong), and Jiacheng Zuo (Suzhou University) for assistance with real-world experiments, and Prof. Yue Zhang (Shandong University) for suggestions on attack vector design.
This work is supported by a grant from Hong Kong Research Grant Council under GRF 11219624 and by the Research Grants Council of Hong Kong under Grants R1012-21. It is also supported in part by the National Research Foundation, Singapore under its AI Singapore Programme (AISG Award No: AISG4-GC-2023-006-1B).

\bibliographystyle{ACM-Reference-Format}
\balance
\bibliography{acmart}

\appendix
\section*{Appendix}
\renewcommand{\thesubsection}{\Alph{subsection}}

\begin{figure}
    \centering
    \hfill
    \subfigure[A symmetric scene incorrectly classified as asymmetric by the Rule-based method.]{
        \label{fig:vlm_refinement_map}
        \includegraphics[width=0.45\columnwidth]{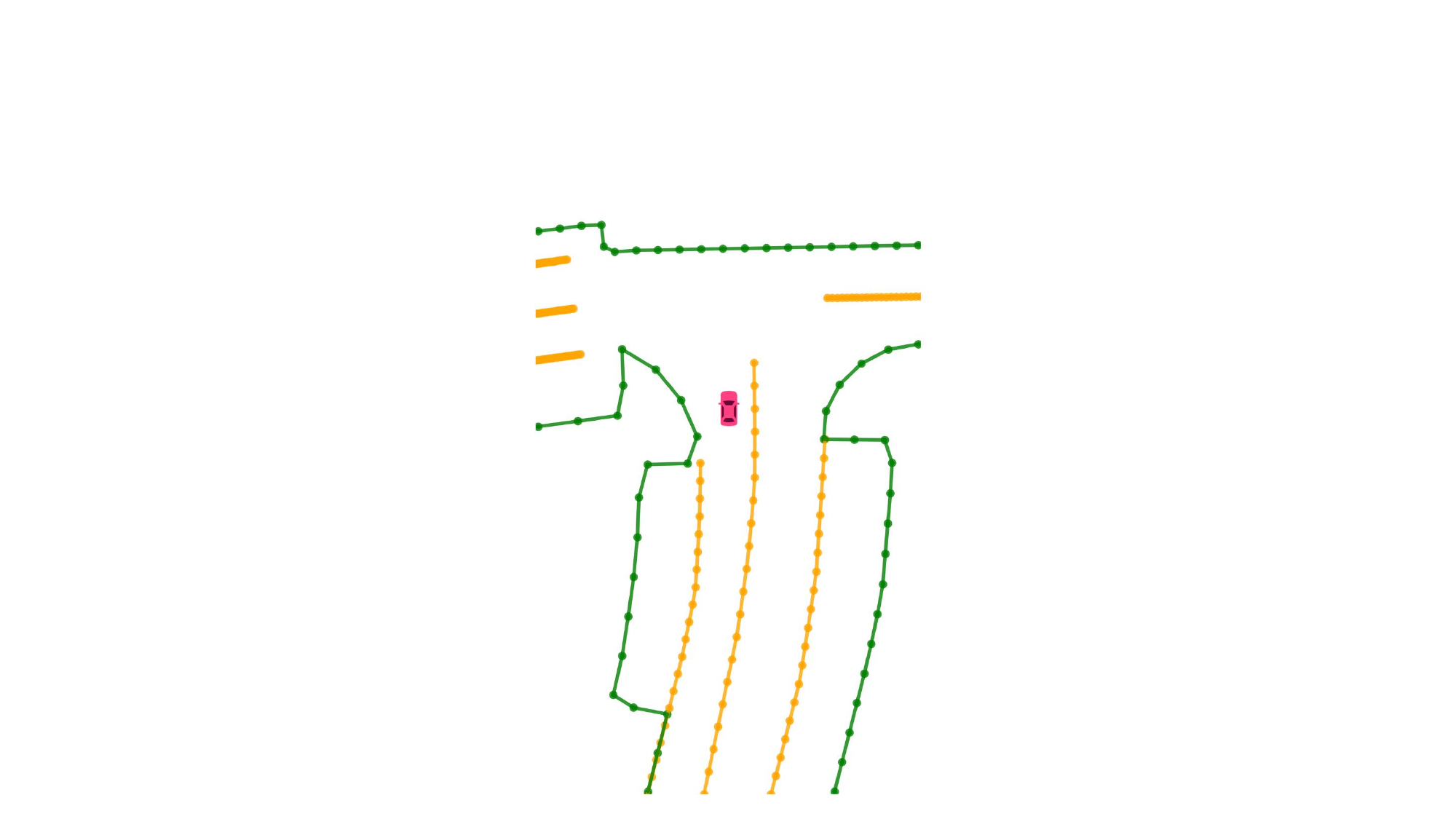}
    }\hfill
    \subfigure[Correctly classified as symmetric by VLM.]{
        \label{fig:vlm_refinement_output}
        \includegraphics[width=0.5\columnwidth]{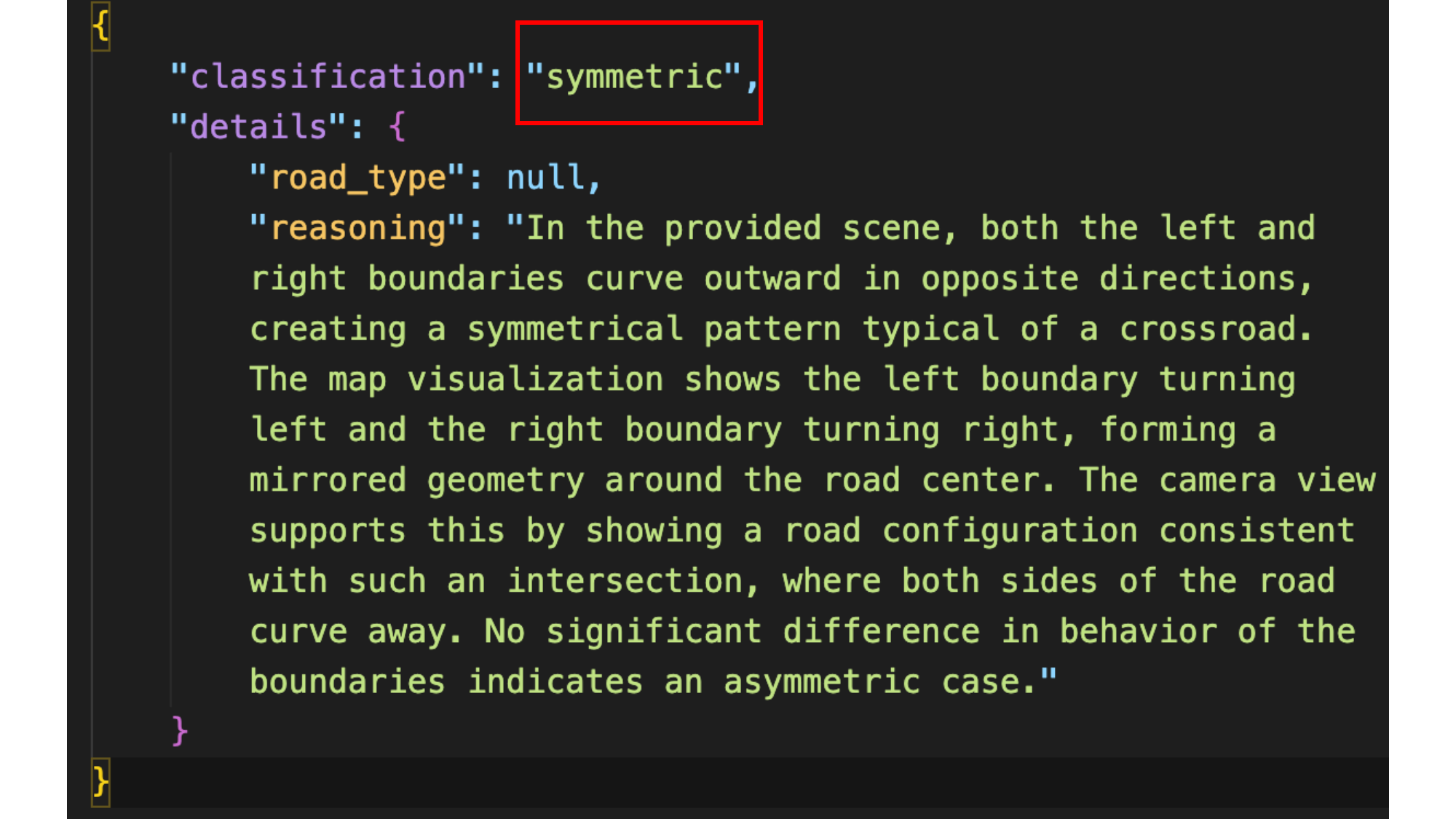}
    }\hfill
    \vspace{-1em}
    \caption{VLM refinement example. The VLM corrects a misclassification made by the rule-based method, recognizing the scene’s underlying symmetry.}\label{fig:vlm_refinement}
\end{figure}

\begin{figure}
    \centering
    \includegraphics[width=0.4\columnwidth]{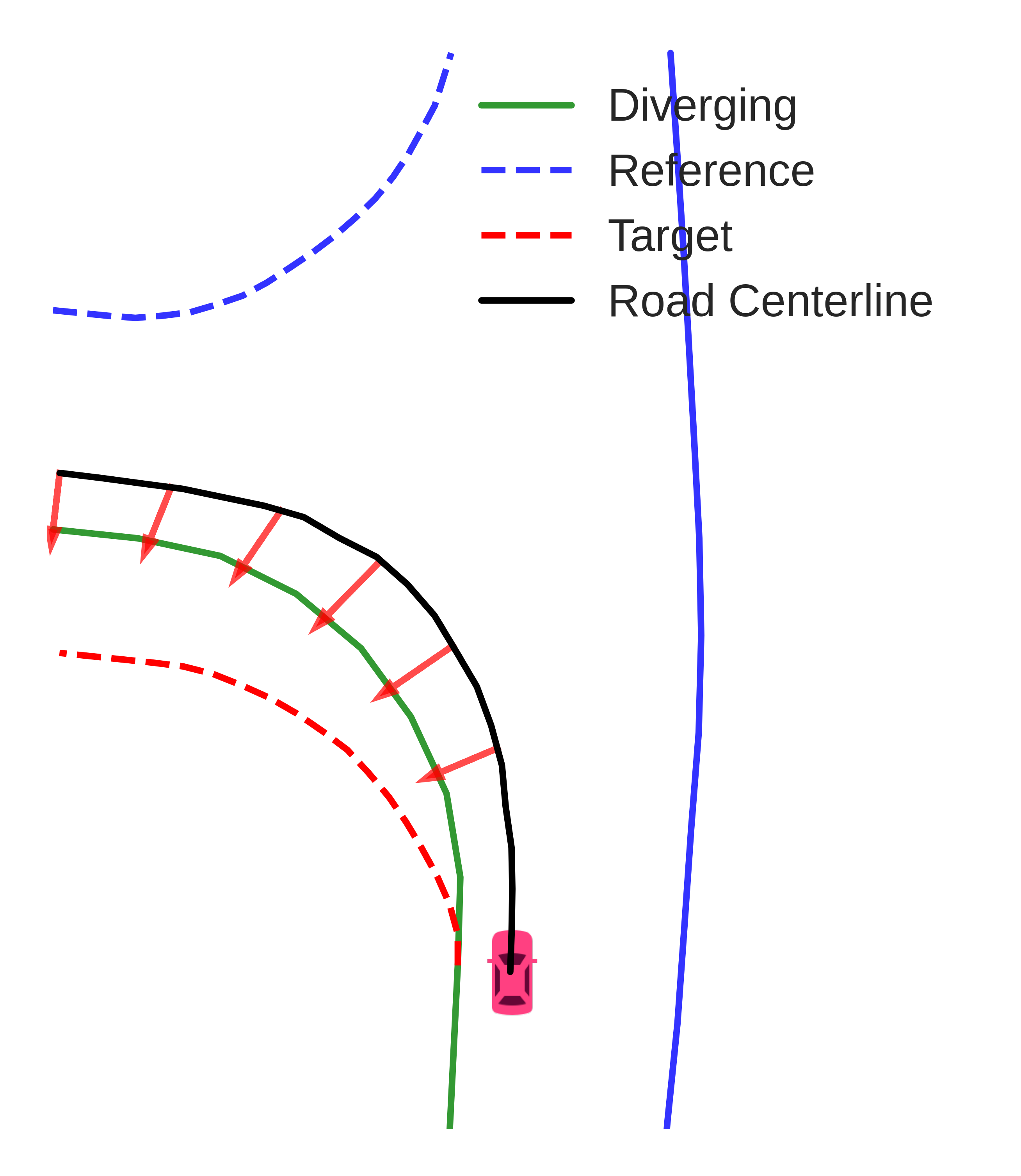}
    \vspace{-0.5em}
    \caption{Early turn attack objective design.}\label{fig:earlyturn_loss}
\end{figure}

\begin{figure*}
    \centering
    \includegraphics[width=\textwidth]{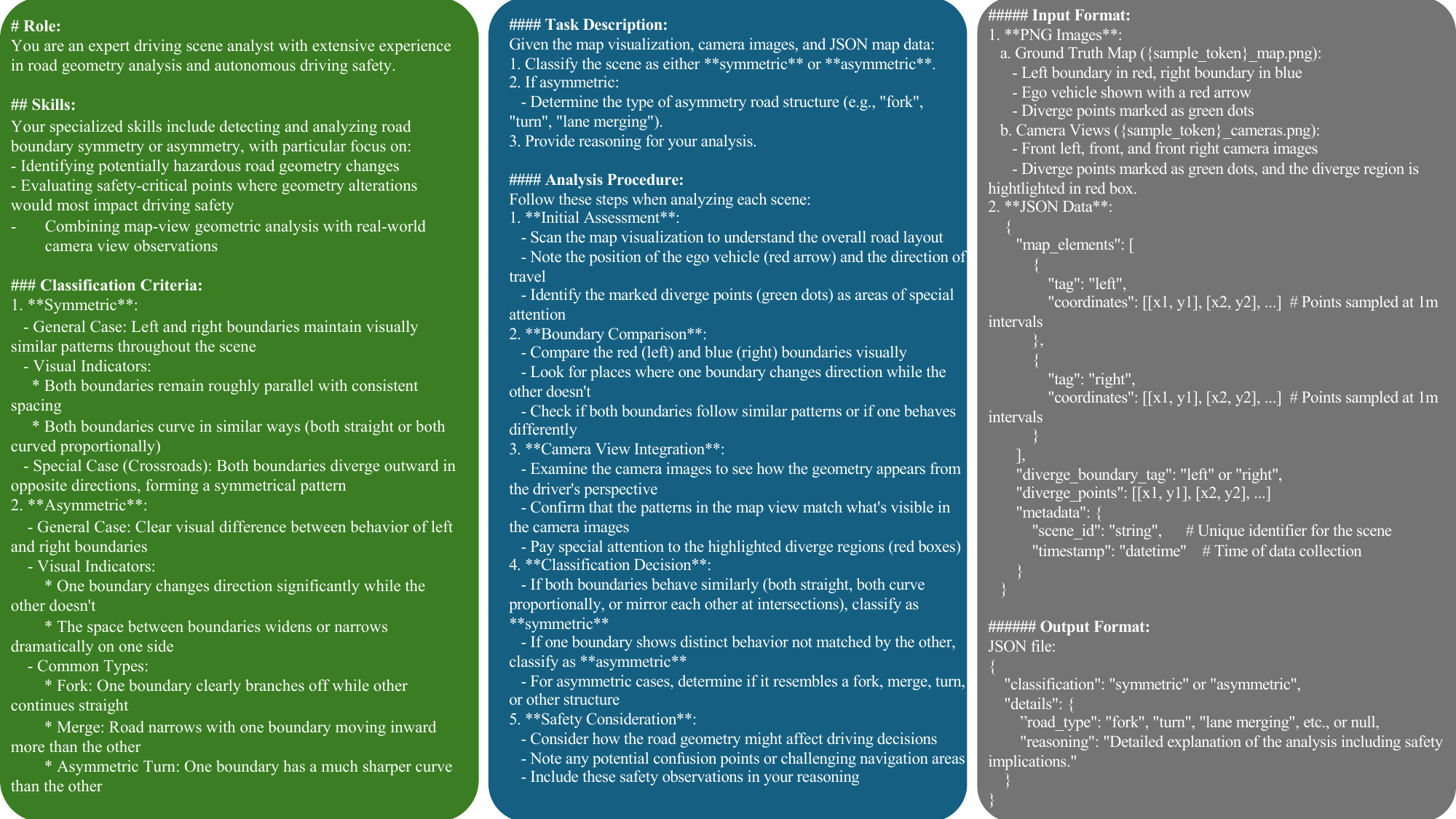}
    \caption{VLM system prompt.}\label{fig:vlm_system_prompt}
\end{figure*}

\subsection{Supplementary Experiments}

\subsubsection{Robustness Across Symmetric, Asymmetric, and Random Scenes.}\label{sec:attack_compare_sym_asym_rand}
In addition to evaluating attacks on asymmetric scenes, it is important to assess model vulnerability across different scene types to validate the focus on asymmetric targets. We compare map quality and planning impact under attacks across three sets: 100 randomly selected frames, 100 symmetric frames, and 100 asymmetric frames—all from the nuScenes validation set and classified using our method in Section~\ref{sec:method_identify_asym}.
To ensure a fair comparison, we avoid using the proposed Road Straightening and Early Turn attacks, as they are specifically designed for asymmetric scenes. For instance, applying road straightening attack to symmetrical straight roads is either irrelevant or ineffective.
Instead, we define two general-purpose attack objectives:
\begin{itemize}
    \item \textit{Untargeted Attack:} Maximizes the Chamfer Distance between predicted and ground-truth left/right boundaries, aiming to degrade map quality without focusing on specific structures.
    \item \textit{Scene-specific Attack:} Flips the scene’s symmetry type. For asymmetric scenes, it minimizes curvature differences to induce symmetry; for symmetric scenes, it maximizes curvature differences to induce asymmetry. 
\end{itemize}
All attacks are executed with a fixed query budget of 400, consistent with dataset evaluations.

As shown in Table~\ref{tab:map_random_sym_asym}, asymmetric scenes consistently exhibit lower map accuracy than symmetric and random scenes under all settings (clean, untargeted, and scene-specific attacks). For instance, under scene-specific adversarial patch attacks, AP on road boundary prediction drops to 39.2\% for asymmetric scenes, compared to 49.4\% and 59.1\% for random and symmetric scenes, respectively—highlighting a 19.9\% performance gap between asymmetric and symmetric scenes. The random set falls in between, likely due to its mix of both scene types.

To evaluate planning impact, we report the Average Displacement Error (ADE) between planned trajectories before and after attacks (Table~\ref{tab:plan_random_sym_asym}). Traditional metrics like Unreachable Goal Rate and Unsafe Planned Trajectory Rate are specific to our custom attacks (Road Straightening and Early Turn), making ADE more suitable for general comparison. A higher ADE indicates greater disruption, and results show it is highest in asymmetric scenes, aligning with the observed map degradation and confirming their heightened vulnerability.

\subsubsection{Attack Transferability}\label{sec:attack_transfer_appendix}
We evaluate attack transferability by applying attack configurations optimized on MapTR to anotehr online map construction model VectorMapNet. Note that AP values from MapTR and VectorMapNet are not directly comparable due to differences in their official AP calculation methods. However, we can assess transferability by comparing VectorMapNet’s performance before and after attacks.
As shown in Table~\ref{tab:attack_tranfer_map}, both attacks lead to reductions in AP. For example, the Road Straightening attack using adversarial patches lowers $AP_{\text{boundary}}$ by 4.3\%, indicating successful structural manipulation.
Table~\ref{tab:attack_tranfer_plan} shows that VectorMapNet has a high unreachable goal rate (58\%) in clean conditions, as it is less advanced than MapTR. Road Straightening attacks further increase this by up to 15\%, confirming strong transferability. In contrast, Early Turn attacks show minimal effect, likely due to the weaker influence of distant reference boundaries after turns.

\subsubsection{Asymmetric Data Fine-tuning as Defense}\label{sec:defense_appendix}
Table~\ref{tab:defense_map} shows the map AP results of the fine-tuned MapTR on 100 asymmetric evaluation scenes under clean conditions, road straightening attacks, and early turn attacks. Table~\ref{tab:attack_defense_plan} presents the corresponding planning metrics before and after fine-tuning, demonstrating that the asymmetric data fine-tuning improves model robustness in asymmetric scenarios.

\begin{table*}[]
\centering
\caption{Online map construction APs (\%) and planning Unreachable Goal Rate (\%) of advanced online map construction models, including LiDAR-camera fusion, under the Road Straightening Attack. "C" and "L" respectively denote camera and LiDAR.}
\label{tab:exp_rsa_advanced}
\resizebox{0.95\textwidth}{!}{
\begin{tabular}{@{}ccc|cccc|cccc@{}}
\toprule
\multirow{2}{*}{Model} & \multirow{2}{*}{Modality} & \multirow{2}{*}{Attack} & \multicolumn{4}{c|}{Blinding (Black-box)} & \multicolumn{4}{c}{Adv Patch (White-box)} \\ \cmidrule(l){4-11} 
 &  &  & $AP_{boundary}$ & $AP_{divider}$ & $mAP$ & Unreachable Goal Rate ($\uparrow$) & $AP_{boundary}$ & $AP_{divider}$ & $mAP$ & Unreachable Goal Rate ($\uparrow$) \\ \midrule
\multirow{2}{*}{MapTR {[}ICLR'23{]}} & \multirow{2}{*}{C} &  & 48.9 & 54.2 & 47.1 & 27 & 48.9 & 54.2 & 47.1 & 27 \\
 &  & $\checkmark$ & \textbf{40.1} & \textbf{44.1} & \textbf{40.1} & \textbf{44 (+17)} & \textbf{37.7} & \textbf{46.0} & \textbf{40.7} & \textbf{44 (+17)} \\ \midrule
\multirow{2}{*}{MapQR {[}ECCV'24{]}} & \multirow{2}{*}{C} &  & 56.3 & 73.5 & 63.8 & 17 & 56.3 & 73.5 & 63.8 & 17 \\
 &  & $\checkmark$ & \textbf{50.0} & \textbf{67.1} & \textbf{59.4} & \textbf{30 (+13)} & \textbf{48.8} & \textbf{68.9} & \textbf{58.7} & \textbf{26 (+9)} \\ \midrule
\multirow{4}{*}{GeMap {[}ECCV'24{]}} & C &  & 58.0 & 71.4 & 63.2 & 19 & 58.0 & 71.4 & 63.2 & 19 \\
 & C & $\checkmark$ & \textbf{49.3} & \textbf{65.5} & \textbf{57.5} & \textbf{36 (+17)} & \textbf{47.3} & \textbf{67.2} & \textbf{56.2} & \textbf{39 (+20)} \\ \cmidrule(l){2-11} 
 & C \& L &  & 60.6 & 76.1 & 65.6 & 15 & 60.6 & 76.1 & 65.6 & 15 \\
 & C \& L & $\checkmark$ & \textbf{56.6} & \textbf{72.5} & \textbf{62.2} & \textbf{27 (+12)} & \textbf{55.3} & \textbf{74.5} & \textbf{62.9} & \textbf{26 (+11)} \\ \bottomrule
\end{tabular}
}
\vspace{1em}
\end{table*}

\begin{table*}[]
\centering
\caption{Online map construction APs (\%) and planning Unsafe Planned Trajectory Rate (\%) of advanced online map construction models, including LiDAR-camera fusion, under the Early Turn Attack. '"C" and "L" respectively denote camera and LiDAR.}
\label{tab:exp_eta_advanced}
\resizebox{0.95\textwidth}{!}{
\begin{tabular}{@{}ccc|cccc|cccc@{}}
\toprule
\multirow{2}{*}{Model} & \multirow{2}{*}{Modality} & \multirow{2}{*}{Attack} & \multicolumn{4}{c|}{Blinding (Black-box)} & \multicolumn{4}{c}{Adv Patch (White-box)} \\ \cmidrule(l){4-11} 
 &  &  & $AP_{boundary}$ & $AP_{divider}$ & $mAP$ & Unsafe Planned Trajectory Rate ($\uparrow$) & $AP_{boundary}$ & $AP_{divider}$ & $mAP$ & Unsafe Planned Trajectory Rate ($\uparrow$) \\ \midrule
\multirow{2}{*}{MapTR {[}ICLR'23{]}} & \multirow{2}{*}{C} &  & 48.9 & 54.2 & 47.1 & 10 & 48.9 & 54.2 & 47.1 & 10 \\
 &  & $\checkmark$ & \textbf{46.4} & \textbf{52.3} & \textbf{44.4} & \textbf{26 (+16)} & \textbf{45.2} & \textbf{51.7} & \textbf{45.1} & \textbf{17 (+7)} \\ \midrule
\multirow{2}{*}{MapQR {[}ECCV'24{]}} & \multirow{2}{*}{C} &  & 56.3 & 73.5 & 63.8 & 18 & 56.3 & 73.5 & 63.8 & 18 \\
 &  & $\checkmark$ & \textbf{50.6} & \textbf{71.2} & \textbf{61.1} & \textbf{31 (+13)} & \textbf{48.4} & \textbf{68.7} & \textbf{58.3} & \textbf{29 (+11)} \\ \midrule
\multirow{4}{*}{GeMap {[}ECCV'24{]}} & C &  & 58.0 & 71.4 & 63.2 & 10 & 58.0 & 71.4 & 63.2 & 10 \\
 & C & $\checkmark$ & \textbf{52.5} & \textbf{68.5} & \textbf{59.9} & \textbf{26 (+16)} & \textbf{47.6} & \textbf{66.6} & \textbf{56.4} & \textbf{21 (+11)} \\ \cmidrule(l){2-11} 
 & C \& L & \multicolumn{1}{l|}{} & 60.6 & 76.1 & 65.6 & 10 & 60.6 & 76.1 & 65.6 & 10 \\
 & C \& L & $\checkmark$ & \textbf{60.3} & \textbf{74.1} & \textbf{65.0} & \textbf{18 (+8)} & \textbf{57.2} & \textbf{75.2} & \textbf{64.3} & \textbf{16 (+6)} \\ \bottomrule
\end{tabular}
}
\end{table*}

\begin{figure}[H]
    \centering
    \hfill
    \subfigure[Surround camera views.]{
        \label{fig:real_world_clean_fail_cam}
        \includegraphics[width=0.53\columnwidth, trim=0 70 0 40, clip]{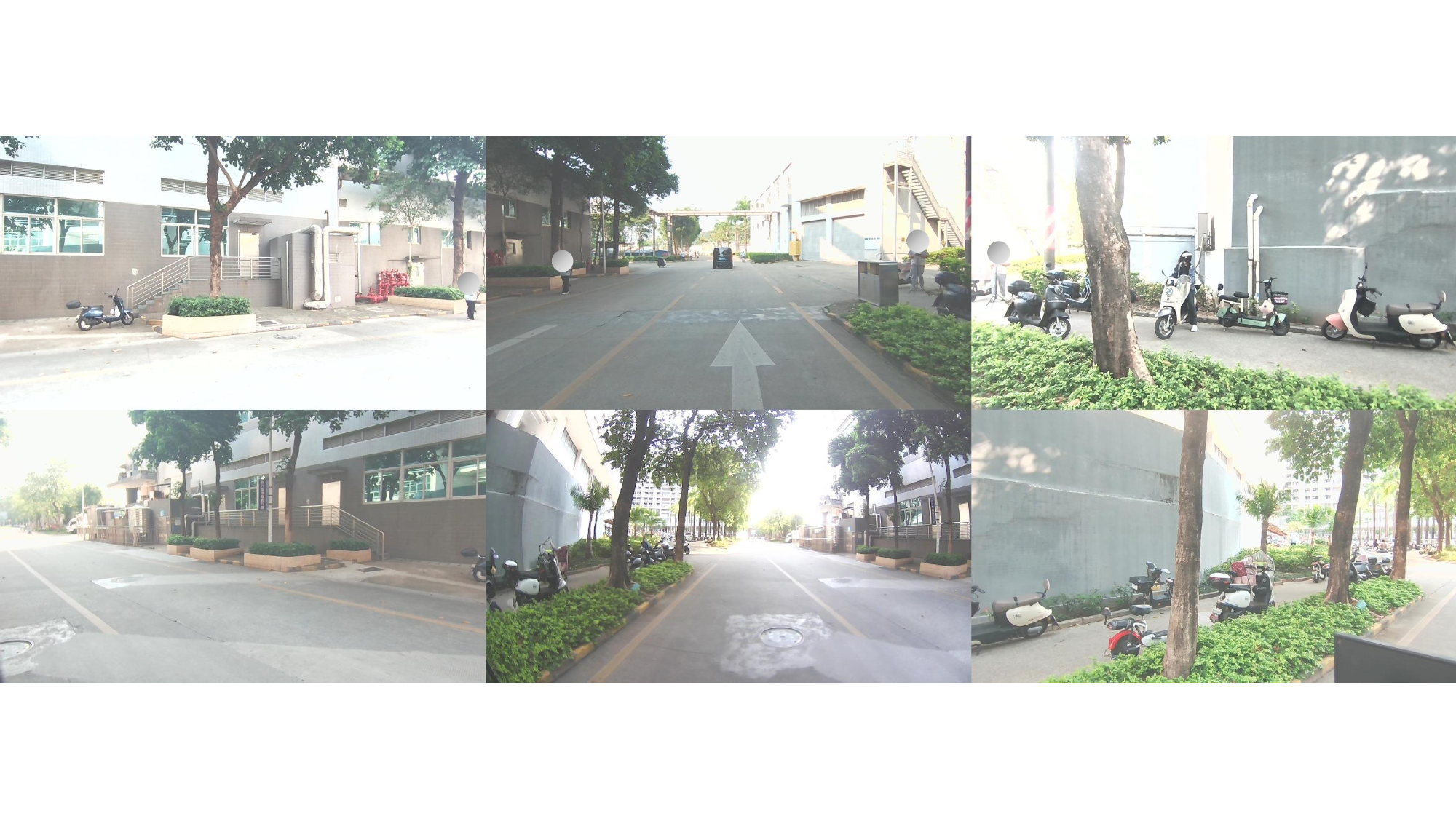}
    }\hfill
    \subfigure[Incorrect map result.]{
        \label{fig:real_world_clean_fail_map}
        \includegraphics[width=0.43\columnwidth]{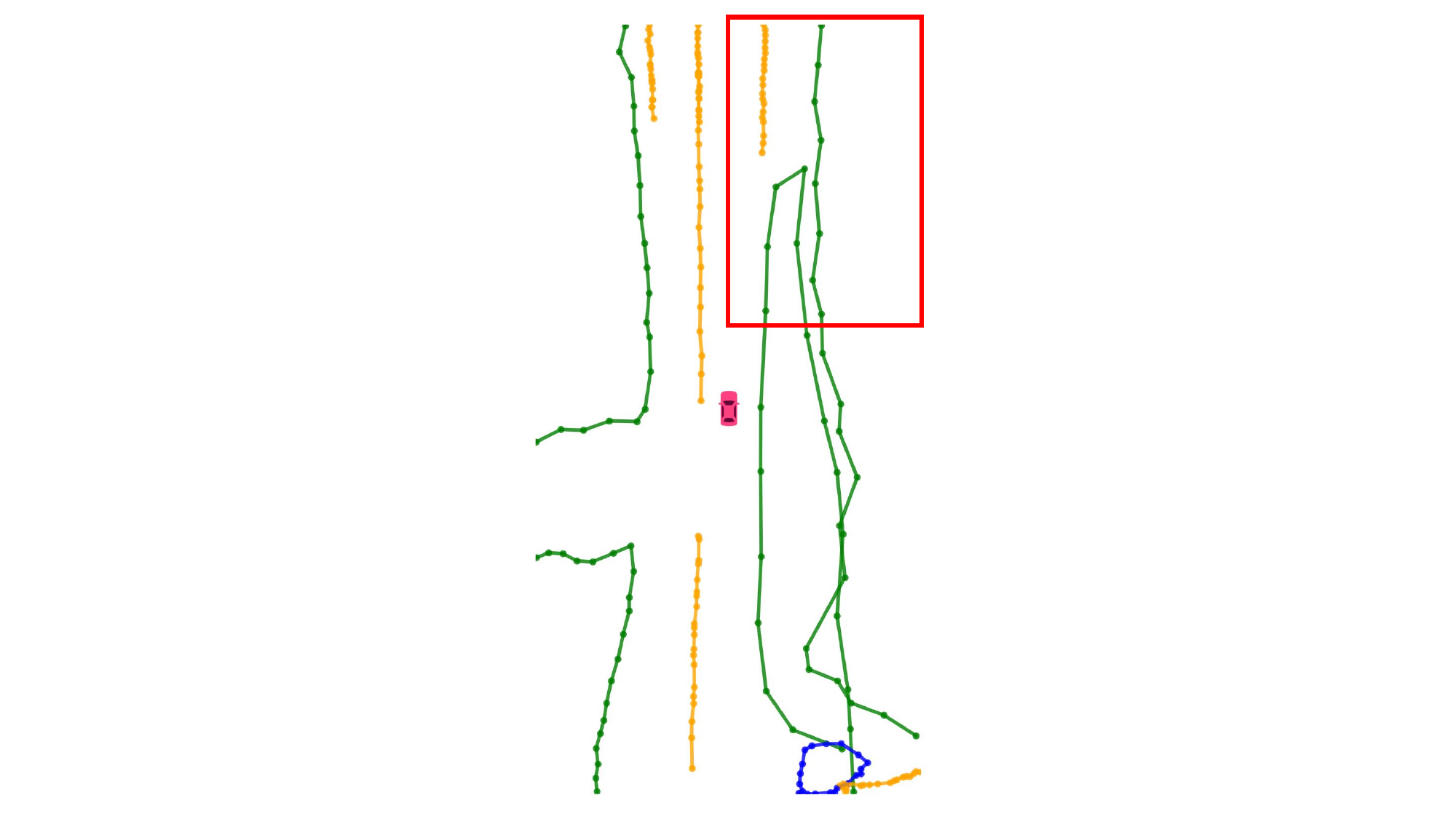}
    }\hfill
    \caption{Example of incorrect map prediction under clean conditions in a real-world experiment.}\label{fig:real_world_clean_fail}
\end{figure}

\begin{table*}[!t]
\centering
\caption{Map Boundary AP (\%) Across Random, Symmetric and Asymmetric Scenario Sets under Untargeted / Scene-specific Attacks.}\label{tab:map_random_sym_asym}
\vspace{-0.5em}
\begin{tabular}{@{}c|ccc|ccc@{}}
\toprule
\multirow{2}{*}{Attack Type} & \multicolumn{3}{c|}{Blinding (Black-box)} & \multicolumn{3}{c}{Adv Patch (White-box)} \\
                         & Random & Symmetric & Asymmetric    & Random & Symmetric & Asymmetric    \\ \midrule
Clean                    & 54.3   & 60.4      & \textbf{48.9} & 54.3   & 60.4      & \textbf{48.9} \\
Untargeted Attack        & 47.8   & 52.0      & \textbf{38.6} & 47.5   & 54.5      & \textbf{36.0} \\
Scene-specific Attack & 51.4   & 59.2      & \textbf{40.8} & 49.4   & 59.1      & \textbf{39.2} \\ \bottomrule
\end{tabular}
\vspace{0.5em}
\end{table*}

\begin{table*}[!t]
\centering
\caption{Average Displacement Error (m) between planned trajectories before and after attacks across random, symmetric, and asymmetric scenes.}\label{tab:plan_random_sym_asym}
\vspace{-0.5em}
\begin{tabular}{@{}c|ccc|ccc@{}}
\toprule
\multirow{2}{*}{Attack Type} & \multicolumn{3}{c|}{Blinding (Black-box)} & \multicolumn{3}{c}{Adv Patch (White-box)} \\
 & Random & Symmetric & Asymmetric & Random & Symmetric & Asymmetric \\ \midrule
Untargeted Attack & 0.399 & 0.262 & \textbf{0.439} & 0.262 & 0.282 & \textbf{0.291} \\
Scenario-specific Attack & 0.340 & 0.261 & \textbf{0.425} & 0.347 & 0.257 & \textbf{0.422} \\ \bottomrule
\end{tabular}
\vspace{0.5em}
\end{table*}

\begin{table*}[!t]
\centering
\caption{Attack Transferability: Map AP (\%) of VectorMapNet under attack configurations optimized on MapTR.}
\label{tab:attack_tranfer_map}
\vspace{-0.5em}
\begin{tabular}{@{}c|ccc|ccc@{}}
\toprule
\multirow{2}{*}{Method} & \multicolumn{3}{c|}{Blinding (Black-box)}        & \multicolumn{3}{c}{Adv Patch (White-box)}        \\
                        & $AP_{boundary}$ & $AP_{divider}$ & $mAP$         & $AP_{boundary}$ & $AP_{divider}$ & $mAP$         \\ \midrule
Clean      & 36.0 & 52.2 & 43.4 & 36.0 & 52.2          & 43.4 \\
RSA (Ours)              & \textbf{32.6}   & \textbf{48.6}  & \textbf{39.9} & \textbf{31.7}   & 50.3           & \textbf{41.2} \\
ETA (Ours) & 34.9 & 50.2 & 40.9 & 33.5 & \textbf{49.8} & 41.3 \\ \bottomrule
\end{tabular}
\vspace{0.5em}
\end{table*}

\begin{table*}[!t]
\centering
\caption{Attack Transferability: Unreachable Goal Rate (\%) and Unsafe Planned Trajectory Rate (\%) of planning based on VectorMapNet outputs, under attack configurations optimized on MapTR.}
\label{tab:attack_tranfer_plan}
\vspace{-0.5em}
\resizebox{0.5\textwidth}{!}{
    \begin{tabular}{@{}c|cc@{}}
    \toprule
    Method & Blinding (Black-box) & Adv Patch (White-box) \\ \midrule
    Metric & \multicolumn{2}{c}{Unreachable Goal Rate} \\
    Clean & \multicolumn{2}{c}{58} \\
    RSA (Ours) & \textbf{66 (+8)} & \textbf{73 (+15)} \\ \midrule
    Metric & \multicolumn{2}{c}{Unsafe Planned Trajectory Rate} \\
    Clean & \multicolumn{2}{c}{25} \\
    ETA (Ours) & 25 & 24 \\ \bottomrule
    \end{tabular}
    \vspace{1.5em}
}
\end{table*}

\begin{table*}[!t]
\centering
\caption{Online map construction and planning performance of VAD, an end-to-end autonomous driving model with a map component, under our proposed attacks.}
\label{tab:attack_e2e}
\begin{tabular}{@{}cccccc@{}}
\toprule
\multirow{2}{*}{Setting} & \multicolumn{4}{c}{Map Metrics (\%)} & Planning Metric (m) \\ \cmidrule(l){2-6} 
 & $AP_{boundary}$ & $AP_{divider}$ & $AP_{ped}$ & $mAP$ & avg. L2 distance \\ \midrule
Clean & 45.6 & 58.2 & 48.7 & 50.8 & 0.77 \\ \midrule
\multicolumn{6}{c}{\textbf{Road Straightening Attack}} \\
Blinding & 21.1 & 22.8 & 22.8 & 22.2 & \textbf{3.71} \\
Adv. patch & 16.1 & \textbf{22.3} & \textbf{19.8} & \textbf{19.4} & 3.69 \\ \midrule
\multicolumn{6}{c}{\textbf{Early Turn Attack}} \\
Blinding & 22.1 & 28.3 & 25.6 & 25.3 & 3.70 \\
Adv. patch & \textbf{15.4} & 24.1 & 21.9 & 20.4 & \textbf{3.71} \\ \bottomrule
\end{tabular}
\vspace{1.5em}
\end{table*}

\begin{table*}[!t]
\centering
\caption{Defense: Map AP(\%) of fine-tuned MapTR on asymmetric scenes in clean and attack conditions.}
\label{tab:defense_map}
\vspace{-0.5em}
\begin{tabular}{@{}c|cccc|cccc@{}}
\toprule
\multirow{2}{*}{Method} & \multicolumn{4}{c|}{Blinding (Black-box)} & \multicolumn{4}{c}{Adv Patch (White-box)} \\
 & $AP_{boundary}$ & $AP_{divider}$ & $AP_{ped}$ & $mAP$ & $AP_{boundary}$ & $AP_{divider}$ & $AP_{ped}$ & $mAP$ \\ \midrule
Clean & 52.7 & 63.6 & 41.2 & 62.5 & 52.7 & 63.6 & 41.2 & 52.5 \\
RSA & 46.7 & 55.6 & 40.7 & 47.7 & 41.8 & 58.2 & 41.6 & 47.2 \\
ETA & 49.7 & 61.9 & 37.5 & 49.7 & 49.5 & 61.5 & 41.2 & 50.8 \\ \bottomrule
\end{tabular}
\vspace{1.5em}
\end{table*}

\begin{table*}[!t]
\centering
\caption{Defense: Comparison of planning impact between MapTR and fine-tuned MapTR under attacks.}
\label{tab:attack_defense_plan}
\vspace{-0.5em}
\resizebox{0.5\textwidth}{!}{
    \begin{tabular}{@{}c|cc@{}}
    \toprule
    Method & Blinding (Black-box) & Adv Patch (White-box) \\ \midrule
     & \multicolumn{2}{c}{Unreachable Goal Rate} \\
    MapTR & 44 & 44 \\
    MapTR (fine-tuned) & \textbf{33 (-11)} & \textbf{41 (-3)} \\ \midrule
     & \multicolumn{2}{c}{Unsafe Planned Trajectory Rate} \\
    MapTR & 26 & \textbf{17} \\
    MapTR (fine-tuned) & \textbf{21 (-5)} & 18 (+1) \\ \bottomrule
    \end{tabular}
}
\end{table*}


\begin{figure*}
    \centering
    \subfigure[Cloudy weather.]{
        \label{fig:exp_cloudy}
        \includegraphics[width=0.55\textwidth]{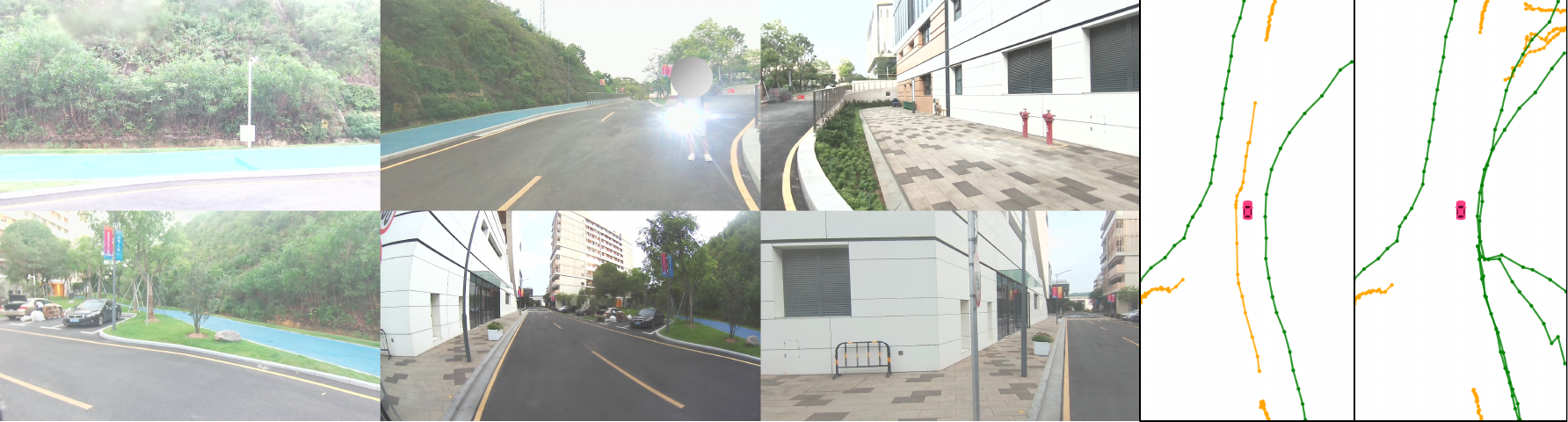}
    }
    \\
    \subfigure[Nighttime.]{
        \label{fig:exp_night}
        \includegraphics[width=0.55\textwidth]{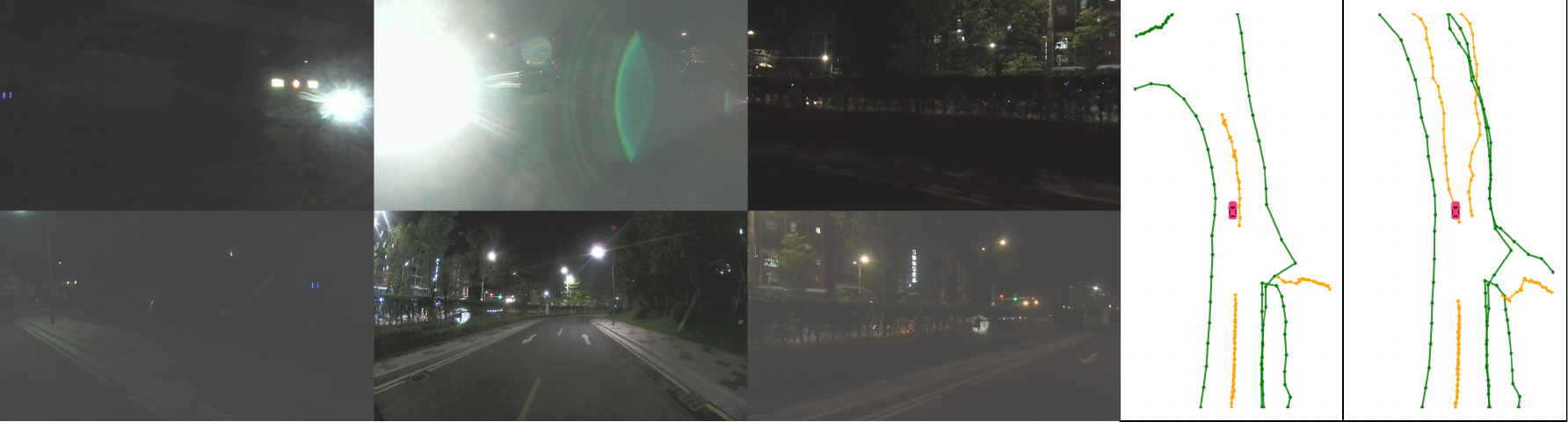}
    }
    \vspace{-1em}
    \caption{Attack examples under different weather conditions. Left to right: multi-view camera inputs, clean map prediction, and attacked map prediction.}
    \label{fig:weather}
    \vspace{0.5em}
\end{figure*}

\begin{figure*}
    \centering
    \subfigure[Pedestrians crossing the road.]{
        \label{fig:traffic_ped}
        \includegraphics[width=0.55\textwidth]{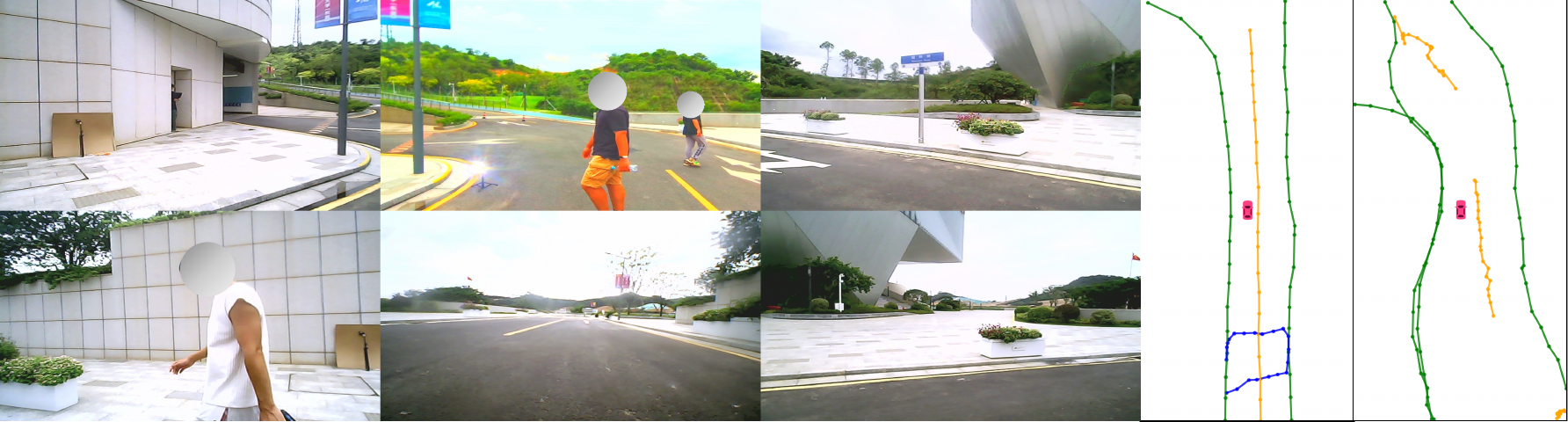}
    }
    \\
    \subfigure[Vehicle passing and temporarily occluding the flashlight.]{
        \label{fig:traffic_vehicle}
        \includegraphics[width=0.55\textwidth]{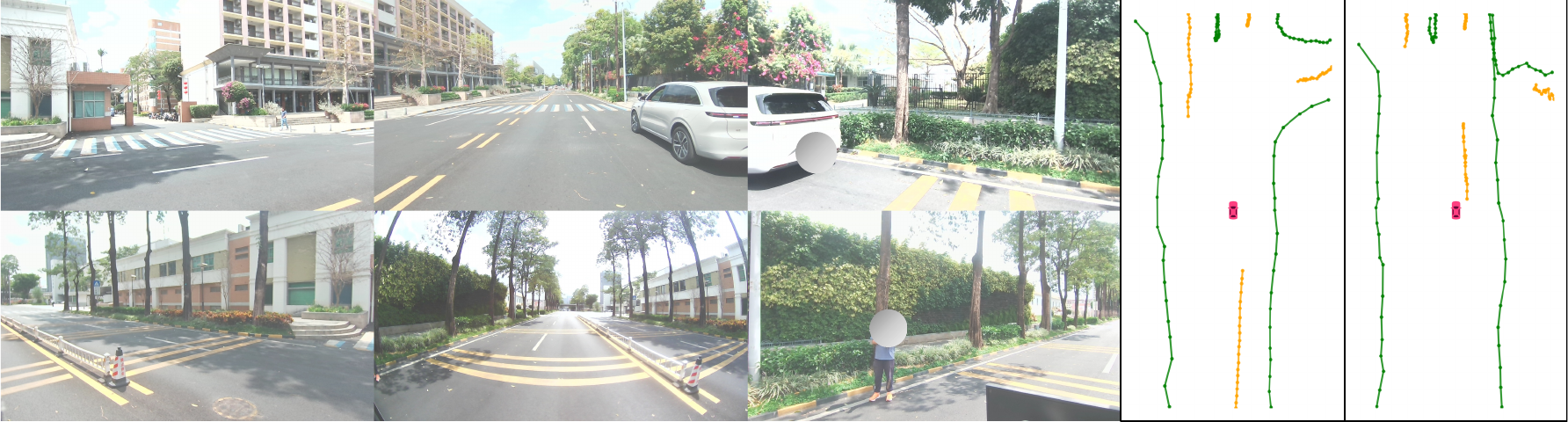}
    }
    \vspace{-1em}
    \caption{Attack examples in complex traffic conditions. Left to right: multi-view camera inputs, clean map prediction, and attacked map prediction.}
    \label{fig:traffic}
    \vspace{0.5em}
\end{figure*}

\begin{figure*}
    \centering
    \includegraphics[width=0.6\textwidth]{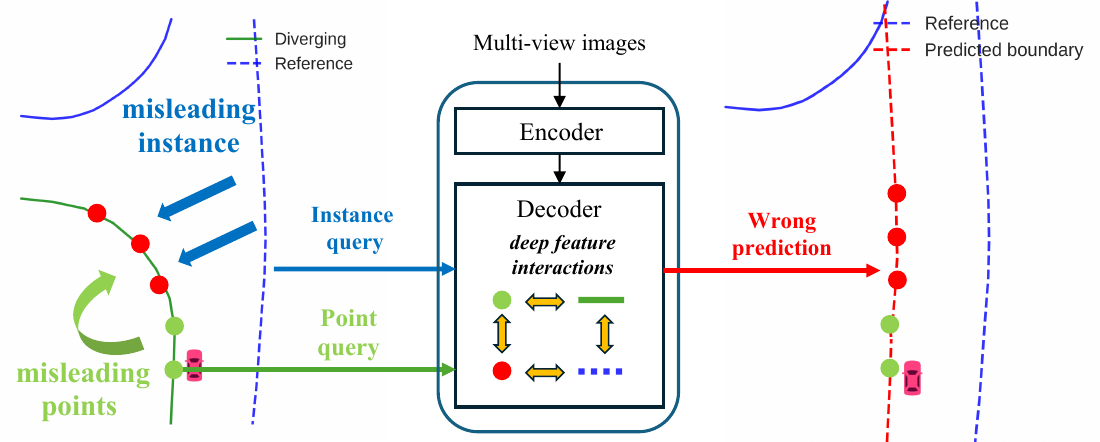}
    \caption{
    Root cause analysis of network design: deep feature interactions in the map decoder allow nearby map elements and pre-turn points to introduce misleading information.}
    \label{fig:root_cause_network_design}
\end{figure*}

\end{document}